\documentclass[fleqn,usenatbib]{mnras}

\usepackage{newtxtext,newtxmath}
\DeclareRobustCommand{\VAN}[3]{#2}
\let\VANthebibliography\thebibliography
\def\thebibliography{\DeclareRobustCommand{\VAN}[3]{##3}\VANthebibliography}

\usepackage[T1]{fontenc}
\usepackage{graphicx}	% Including figure files
\graphicspath{ {./Figures/} }
\usepackage{amsmath}	% Advanced maths commands
\usepackage{multirow}
\usepackage{pdflscape}
\usepackage{longtable}
\usepackage{caption}
\usepackage{anyfontsize}    % Fix for LaTeX error

\defcitealias{GaiaDR3}{\textit{Gaia} DR3}

\usepackage[utf8]{inputenc}
\usepackage{xspace}
\usepackage{verbatim}

\date{October 2022}
\title[TESS phase modulation binaries]{A catalog of binary stars from phase modulation in the first four years of \textit{TESS} Mission photometry}

\author[]{Shishir Dholakia,$^1$
Simon J. Murphy,$^1$ 
Chelsea X. Huang,$^1$
Alexander Venner,$^1$
Duncan Wright$^1$
\\
$^{1}$University of Southern Queensland, Centre for Astrophysics, West Street, Toowoomba, QLD 4350 Australia\\
}
\begin{document}

\maketitle

\begin{abstract}
{We present a catalog of binary companions to $\delta$\,Scuti stars, detected through phase modulations of their pulsations in \textit{TESS} data. Pulsation timing has provided orbits for hundreds of pulsating stars in binaries from space-based photometry. We have applied this technique to $\delta$\,Sct stars observed in the first four years of \textit{TESS} Mission photometry. We searched the 2-min cadence light curves of 1161 short-period instability strip pulsators for variations in pulsation phase caused by the dynamical influence of an unseen companion. We discovered 53 new binaries and we present orbital parameters and mass functions for the 24 systems with solvable orbits. For the brightest star in our sample $\alpha$ Pictoris, we perform a joint fit of the pulsation timing and \textit{Hipparcos} astrometry. We present the first orbit for the $\alpha$ Pictoris system, obtaining an orbital period of 1316$\pm$2 days and a mass for $\alpha$ Pic b of 1.05$\pm$0.05 M$_\odot$. We revisit pulsation timing binaries from \textit{Kepler} with \textit{Gaia} kinematics, finding four systems that are members of the Galatic thick disk or halo. This suggests that they have been rejuvenated by mass transfer, and their companions are now white dwarfs. Further follow up of these systems may yield valuable constraints of the galactic blue straggler population.}
\end{abstract}

\section{Introduction}
Binary stars outnumber single stars in our galaxy. Their prevalence, evolution, and demography are therefore central to understanding galactic and stellar astrophysics. Binary stars with intermediate separations ($\approx$ 50 to 1000 day orbital periods) are particularly astrophysically valuable; such binaries are linked to important unknowns of stellar physics, such as the origin of chemical peculiarity \citep{preston1974} and the rotation distribution \citep{meibom2007, Smith2024}. A complete understanding of the origin of blue stragglers and some cataclysmic variables is also limited by the lack of characterizable binary progenitors, many of which are thought to also be intermediate-period \citep{chen2008,maoz2014}. Stellar binaries in which one component of the system can be aged can be leveraged to age and understand any companions \citep{kraus2009}, thereby enabling age constraints of stars that are otherwise challenging to age. These systems, including Sirius-like systems \citep{holberg2013}, intermediate-separation circularized systems \citep{mathieu1988}, and systems with pulsating components that can be asteroseismically aged, therefore have significant astrophysical utility.

Pulsational phase modulation, also known as pulsation timing, has proven an invaluable technique to detect unseen companions to stars, particularly in space-based photometry provided by the \textit{Kepler} mission \citep{murphy2018}.  Phase modulation is best used on coherent pulsators such as $\delta$\,Sct stars, as opposed to stochastically driven pulsators \citep{compton2016}. For these stars, pulsations can be leveraged as an astrophysical clock.  This “pulsation clock” tracks changes in light arrival time due to radial motion of the pulsator along the line of sight due to the gravitational influence of an unresolved binary companion. As the pulsator orbits the barycentre, its pulsations arrive early or late. The resulting phase modulations allow orbital parameters such as the mass function, eccentricity, and orbital period to be determined \citep{murphy2014, murphy2015}.

Binaries detected from this technique are difficult to detect through other means. Phase modulation is more sensitive to longer period binaries, scaling as $\tau \propto P^{2/3}$, where $\tau$ is the amplitude of the time delays and P is the orbital period \citep{murphy2016}. $\delta$\,Sct stars are hot and often rotate rapidly, making detection of companions through radial velocity variations difficult or impossible. Phase modulation has enabled studies of otherwise inaccessible binary demographics, such as the brown dwarf desert and long-period post-mass-transfer binaries \citep{murphy2018}. As such, binaries detected from phase modulation populate an important and under-represented region of stellar binary parameter space at separations between those detected from eclipses and radial velocities and those that are resolvable directly or interferometrically. 

The Transiting Exoplanet Survey Satellite (\textit{TESS}) \citep{Ricker2014} has provided high-quality photometric monitoring of most of the sky as of 2024. The \textit{TESS} mission survey enables access to diverse stellar populations. \textit{TESS} also samples brighter and more nearby targets than \textit{Kepler}, enabling further follow up and more detailed study of the solar neighborhood. \textit{TESS} therefore provides a high-quality dataset with a large sample of bright and characterizable $\delta$\,Sct stars in a variety of stellar populations, including stars in the thick disk/halo and in stellar associations \citep{gaulme2019}. These $\delta$\,Sct stars can be used for phase modulation analysis as was done extensively with \textit{Kepler}. 

\textit{TESS} photometry has already been used to compile nearly all-sky catalogs of $\delta$\,Sct stars and study their properties (\citealt{gootkin2024}; \citealt{barac2022}; \citealt{balona2020}). Many of these stars lie close to the \textit{TESS} Northern and Southern continuous viewing zones, where the baseline and duty cycle of observation is long enough to observe and adequately sample a full orbital period of any intermediate-period companions. In particular, the $\delta$\,Sct star and well-known planet host $\beta$~Pictoris has been searched for phase modulations to attempt to identify the signal of the planetary companions \citep{zieba2024}. Past phase modulation surveys in \textit{TESS} have also concentrated on substellar companions \citep{vaulato2022}, but a systematic search for stellar companions has not yet been undertaken. 
In this paper, we present a search for companions to stars from pulsation modulation in \textit{TESS} photometry. We describe our methodology in Section \ref{sec:id_binary}, identifying pulsators suitable for analysis in Sec. \ref{sec:id_pulsators} and then detailing our approach to searching for pulsation timing variations and our fitting procedure in Sec. \ref{sec:phasemod}. We then test the robustness of our approach with an exercise in Sec. \ref{sec:ir}. Properties of the binaries we identify are presented in Sec. \ref{sec:binary_sample}, with Sec. \ref{sec:alphapic} devoted to results from the target of interest $\alpha$ Pictoris. Sec. \ref{sec:kinematics} presents results from the kinematics of the \textit{TESS} binaries and \textit{Kepler} binary kinematics from \citet{murphy2018}. We summarize and conclude in Section \ref{sec:concl}.

\section{Identification of Binaries}
\label{sec:id_binary}

Here we describe our approach to identifying pulsators suitable for analysis in the \textit{TESS} photometry, identifying phase modulations in that sample of pulsators, and our procedure to fit those phase modulations to obtain orbital parameters for the binaries.

\subsection{Identification of Pulsators}
\label{sec:id_pulsators}
%figure lists:
% (1) histogram with number of sectors
% (2) sky plots showing the location of your targets 
% (3) showing targets in velocity space
% (4) a few nice lomb scargle figures 

 For our analysis, we use 2-min cadence \textit{TESS} photometry up to the end of TESS year 4 observations (Sector 55) produced by the Science Processing Operations Center (SPOC; \citealt{jenkins2016}) pipeline. We construct a target list of stars observed with TESS target pixel stamps with a $T_{\rm eff}$ between 6000\,K and 10,000\,K in the \textit{TESS} Input Catalog (TIC; \citealt{stassun2019}). We select this temperature range slightly beyond the boundaries of the canonical instability strip to account for the bias that possible unresolved binarity can introduce to the estimates of $T_{\rm eff}$. 
 Based on our findings in Sec. \ref{sec:ir}, we only present analyses for stars with $\geq 7$ sectors of 2-min cadence observations to ensure sufficient orbital phase coverage in our solutions. We use the Pre-search Data Conditioning Simple Aperture Photometry (PDCSAP; \citealt{Stumpe2012, Stumpe2014, Smith2012}) data for our phase modulation analysis, excluding poor-quality data by filtering out any cadences having non-zero `QUALITY' flag values.

\begin{figure}
    \includegraphics[width = \hsize]{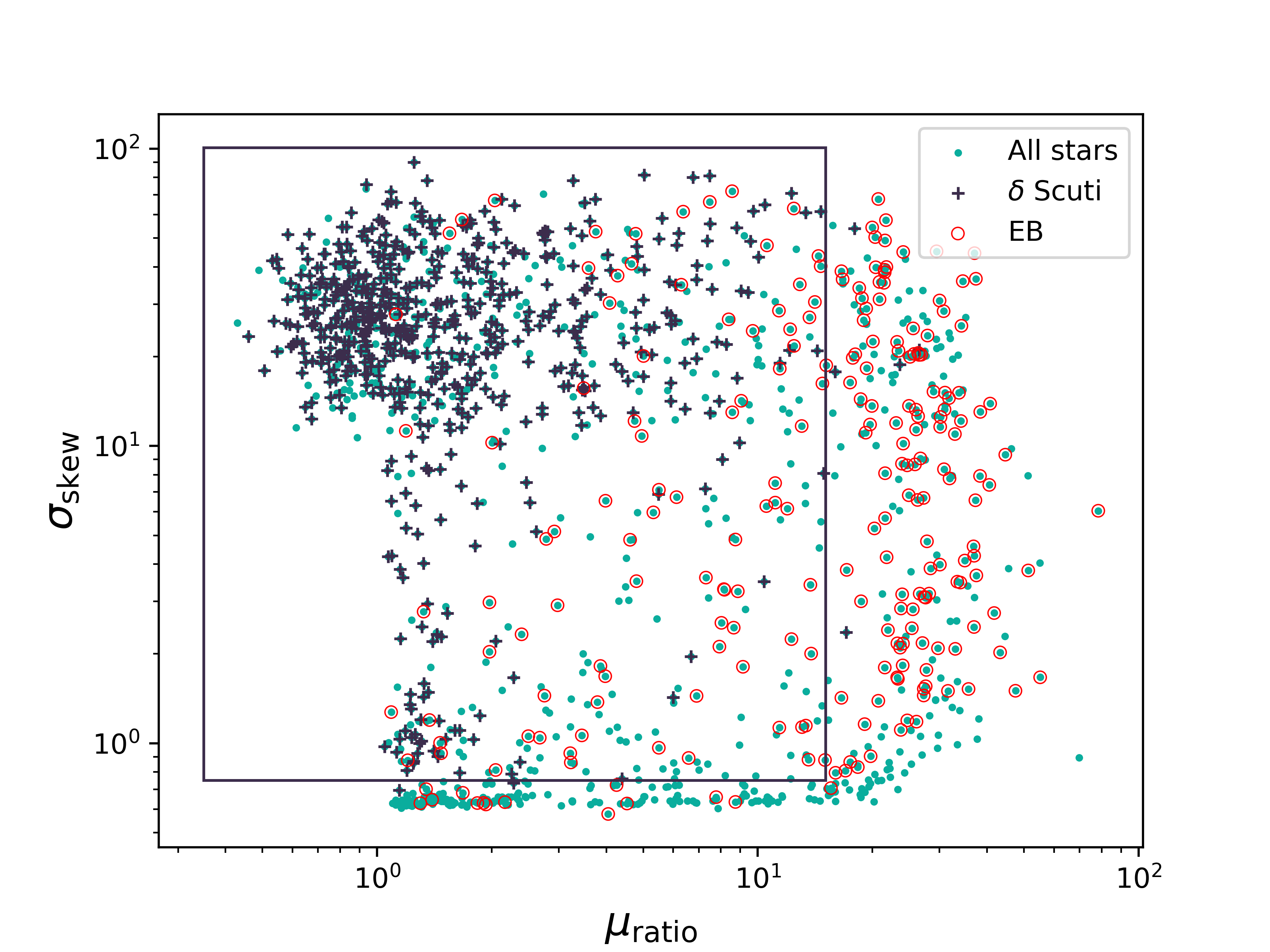}
    \caption{Plot of parameter space used to identify our $\delta$\,Sct stars. The y-axis represents the skew of the periodogram above 10\,d$^{-1}$, and the x-axis plots the median power of the periodogram below 10\,d$^{-1}$ divided by the median power of the periodogram above 10\,d$^{-1}$. These metrics were selected empirically to obtain the best possible separation in our sample between eclipsing binaries and true $\delta$\,Sct stars. Points in cyan represent the full sample of stars. The dark blue plus-markers represent $\delta$\,Sct stars in our human-vetted subsample, whereas red circles represent those vetted as eclipsing binaries. The rectangular region shows the thresholds we use to select our final sample of $\delta$\,Sct stars.}
    \label{fig:skew}
\end{figure}

We calculate a Lomb-Scargle periodogram \citep{lomb1976, Scargle1982} for each light curve with all sectors stitched together. To identify all stars exhibiting periodic behavior, peaks are identified in the periodogram between frequencies of $10$\,d$^{-1}$ and $70$\,d$^{-1}$ if they are above 20 standard deviations from the mean of the periodogram. This significance requirement was selected by trial and error to ensure that modes we use contribute more signal to the phase timing than noise. Stars without any significant peaks identified in this range are discarded from our sample.

To distinguish short-period eclipsing binaries in the sample from $\delta$\,Sct stars, we randomly select a subsample of 500 stars from our full sample. We visually inspect the photometry and periodograms for stars in this subsample, categorizing each by eye as a $\delta$\,Sct star, eclipsing binary, or other. We note characteristics of the periodograms for each category. Based on these observations, we define two heuristics. We first define $\sigma_{\rm skew}$, the skewness of the periodogram  $> 10$\,d$^{-1}$. We find that short-period eclipsing binaries tend to exhibit low skew below a frequency of $10$\,d$^{-1}$ because their orbital periods result in largely monoperiodic variability with rapidly decaying overtones, whereas $\delta$\,Sct stars more often have modes excited throughout their frequency envelopes \citep{uytterhoeven2011}. Secondly, we define $\mu_{\rm ratio}$, the ratio of the means of the Lomb Scargle periodogram above and below $ 10$\,d$^{-1}$. We observe that short-period overcontact binaries tend to exhibit most of their periodogram peaks $< 10$\,d$^{-1}$, making $\mu_{\rm ratio}$ able to disambiguate between these and $\delta$\,Sct stars successfully.

We then create a cut in the parameter space to maximize the inclusion of $\delta$\,Sct stars while limiting the number of eclipsing binaries where possible. This parameter space cut of $\sigma_{\rm skew} > 0.75$ and $\mu_{\rm ratio} < 15$ is displayed in Figure \ref{fig:skew}. From the aforementioned subsample, we note that the fraction of $\delta$\,Sct stars captured in this constraint is ~$99$\%. The fraction of stars in this parameter space that are not $\delta$\,Sct stars is ~$20$\%.

\begin{figure*}
    \includegraphics[width = 0.4\hsize]{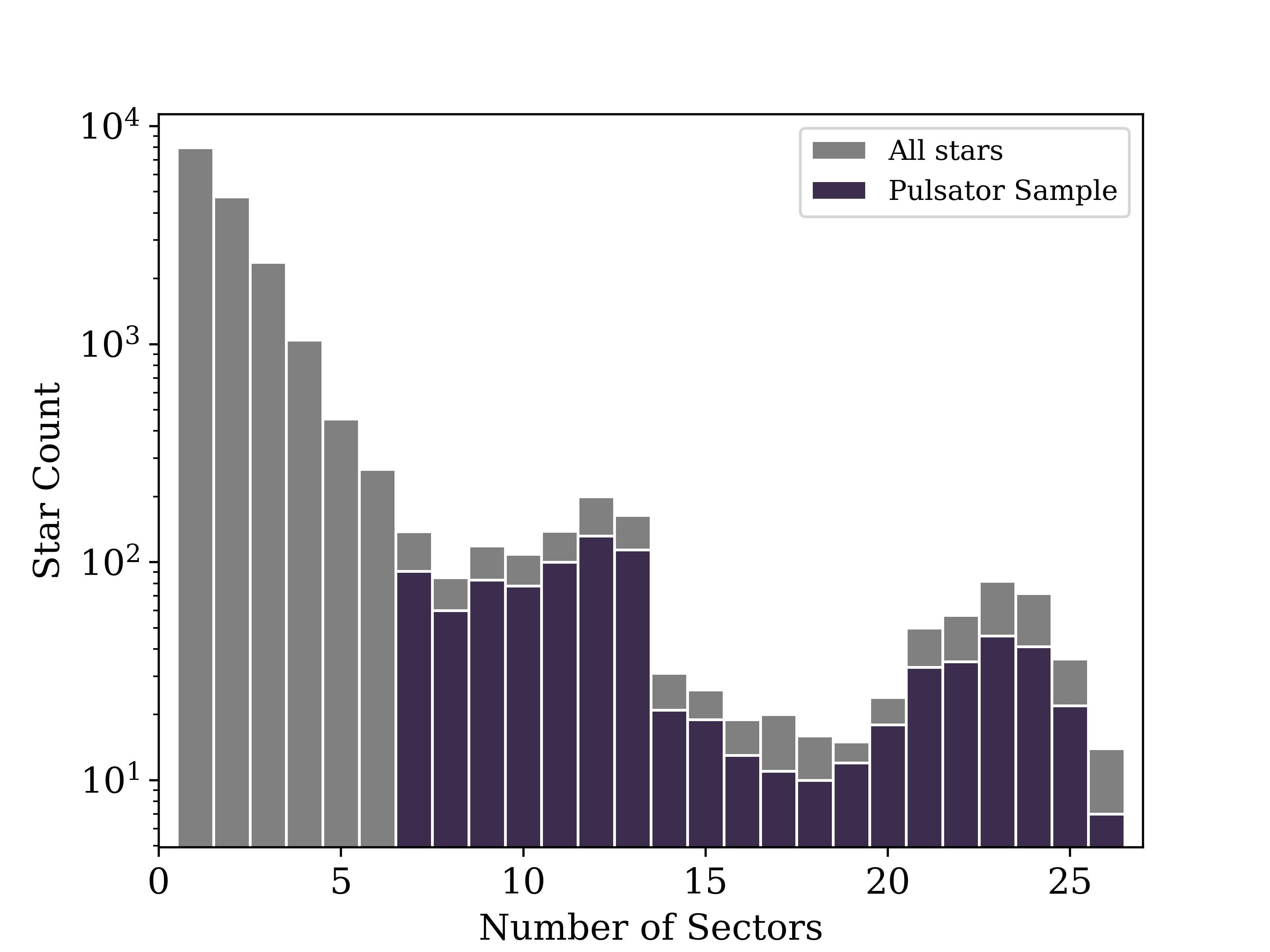}
    \includegraphics[width = 0.4\hsize]{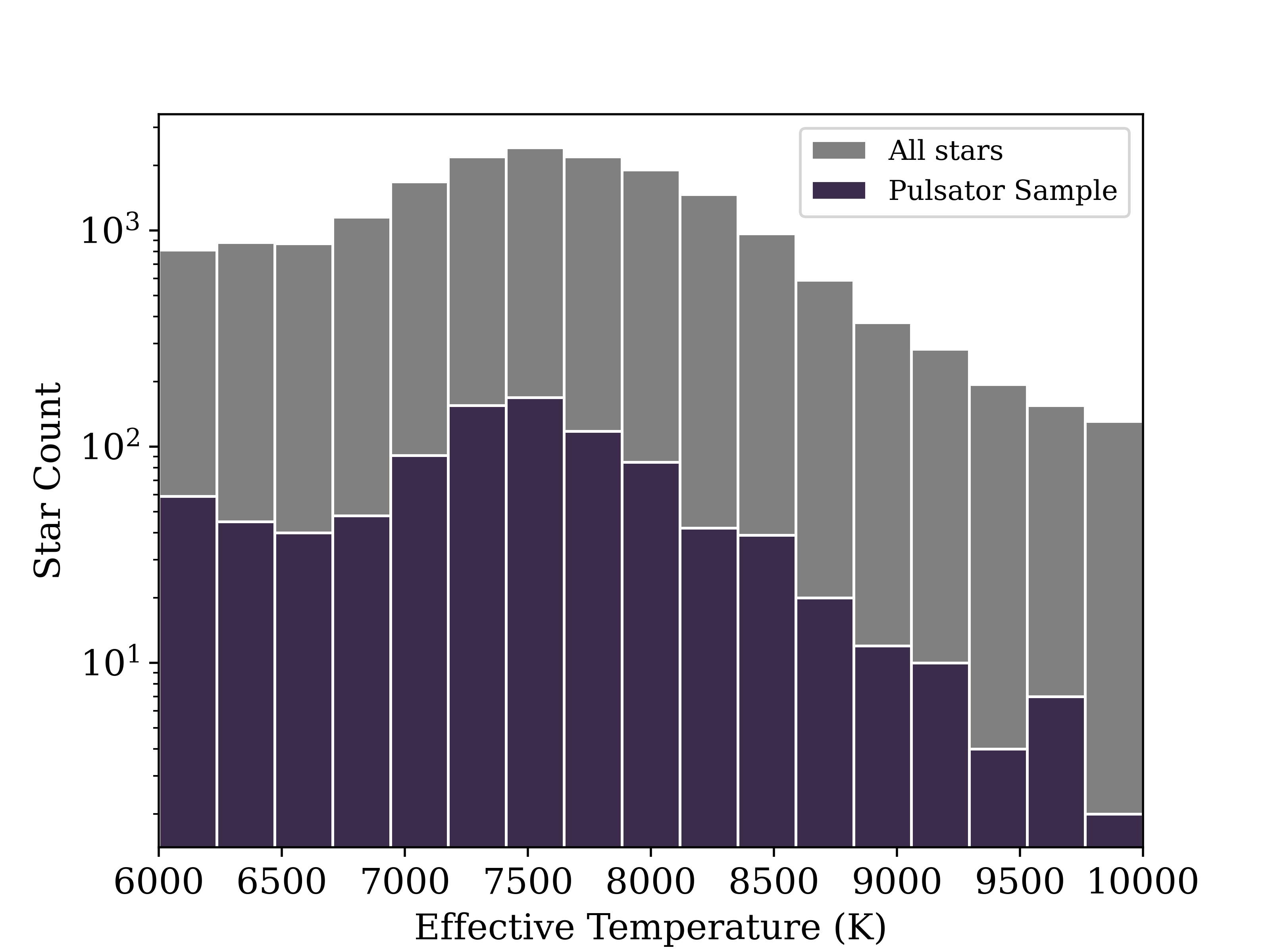}
    \includegraphics[width = 0.4\hsize]{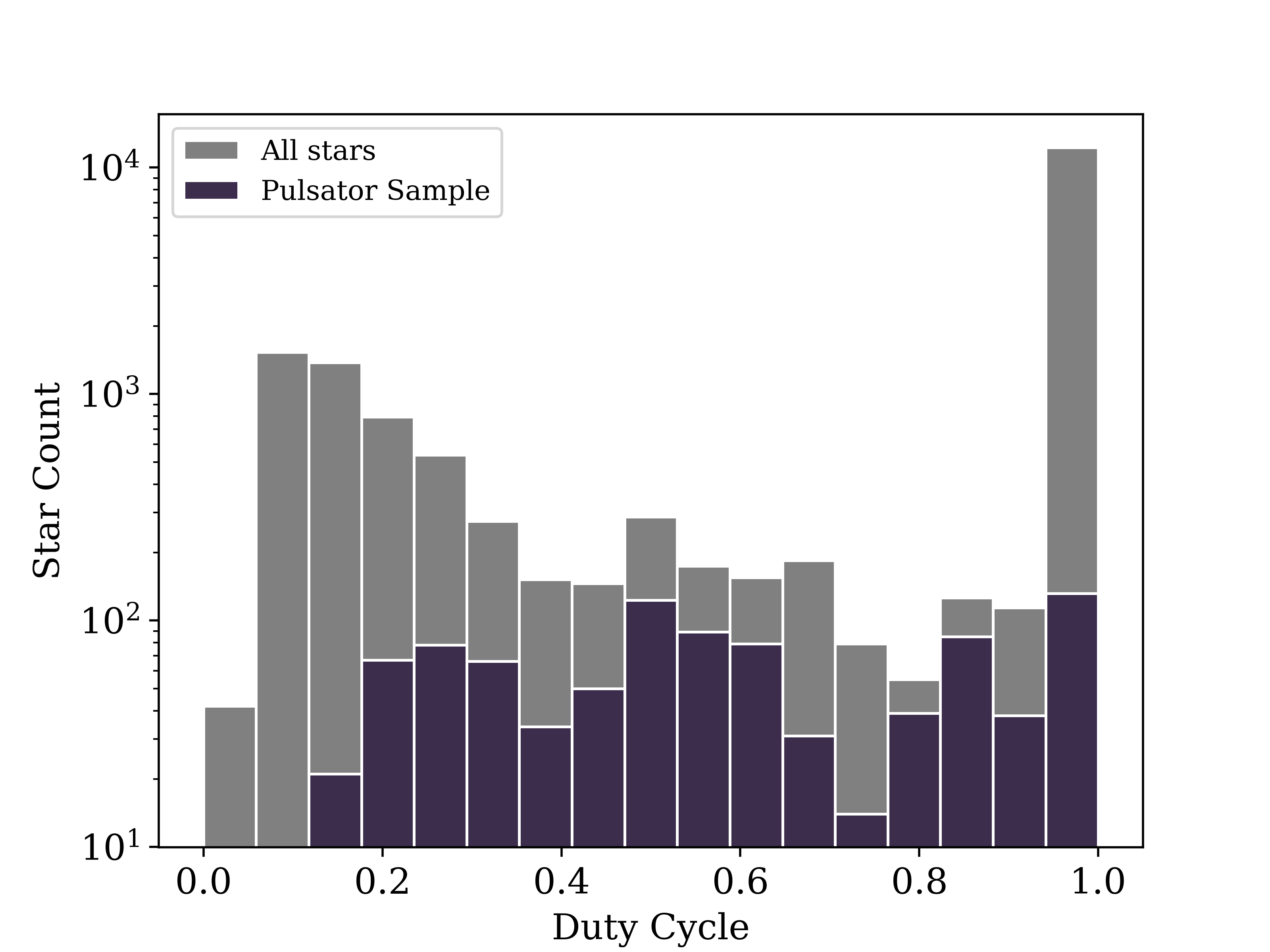}
    \includegraphics[width = 0.4\hsize]{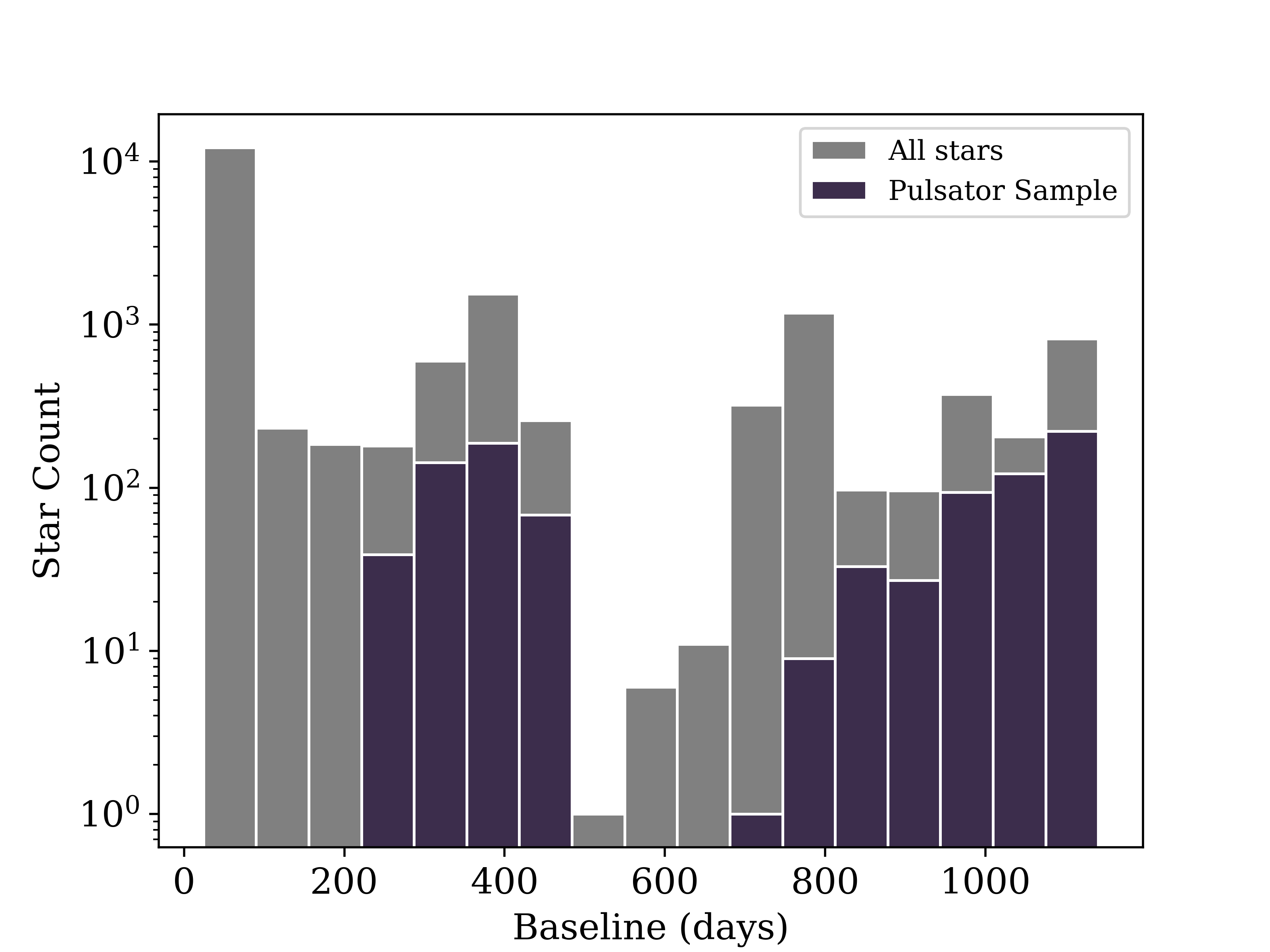}

    \caption{First: The number of sectors of \textit{TESS} photometry available for each target. Second: the distribution of effective temperatures for our targets. Third: Histogram of the duty cycle of \textit{TESS} observations for each target. Fourth: The total baseline of \textit{TESS} observations for each target available up to and including Sector 55. In all histograms, the 1166 $\delta$\,Sct pulsators in our final sample are displayed in purple and all stars in the $T_{\rm eff}$ range are in gray.}
    \label{fig:sectorteffhists}

\end{figure*}

There are a total of 18\,285 stars in the aforementioned $T_{\rm eff}$ range with 2 minute cadence \textit{TESS} data available. Out of these, 1416 stars have $\geq 7$ sectors of \textit{TESS} data available (top panel in Figure \ref{fig:sectorteffhists}). We distill this sample to 1166 variable stars that lie within our parameter space cut -- this is our sample of $\delta$\,Sct stars suitable for analysis. This sample is not volume limited, being biased by the selection effects including the Malmquist bias and the human bias introduced in the targets that are proposed for 2 minute cadence data in \textit{TESS}. Within our sample of pulsators, the distribution of effective temperatures peaks at around 7500\,K and is left-skewed (second panel in Fig.\,\ref{fig:sectorteffhists}).

\subsection{Phase Modulations}
\label{sec:phasemod}

\begin{figure}
    \includegraphics[width = \hsize]{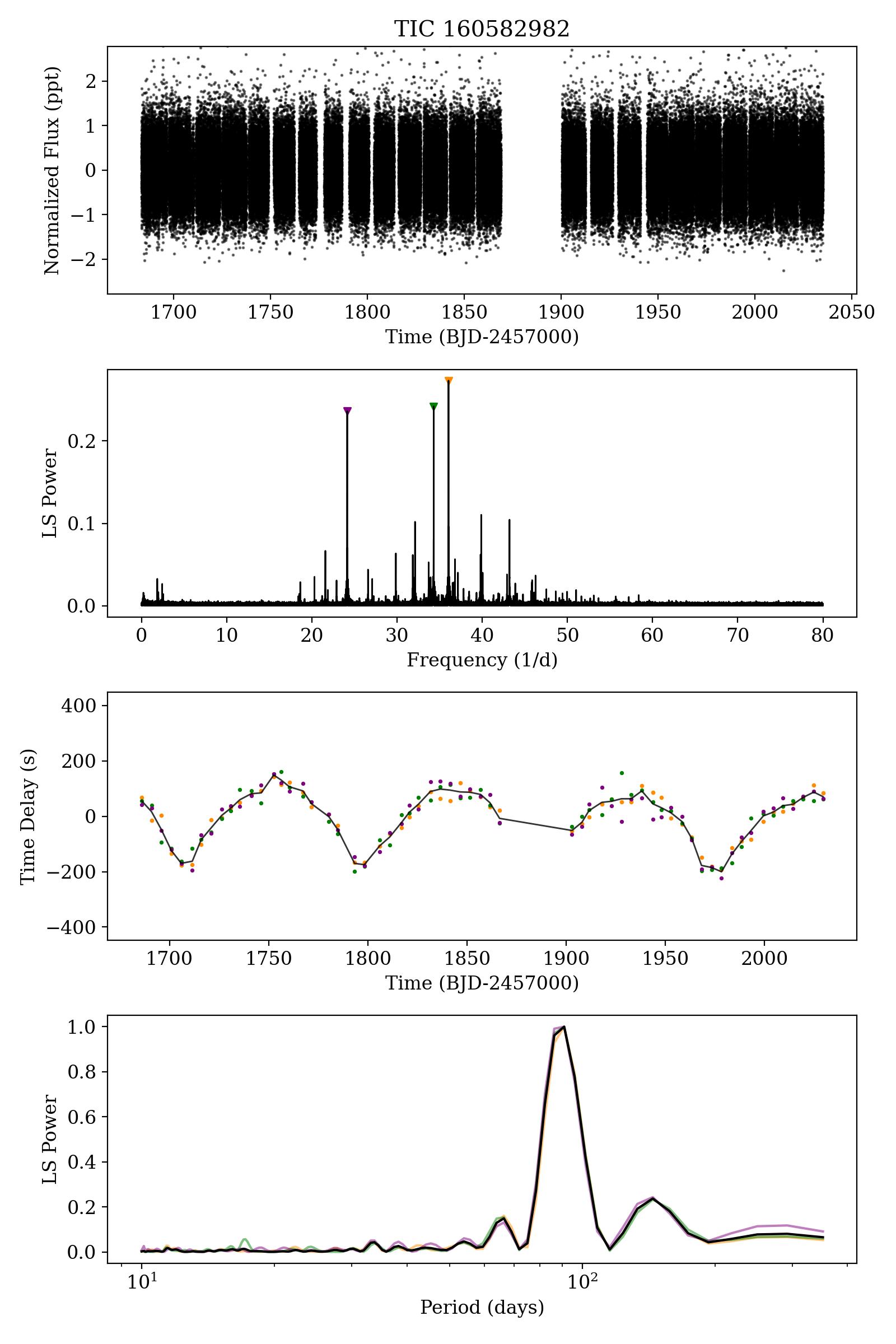}
    \caption{Plots used to shortlist probable pulsation modulation binaries. Top to bottom: Panel~1 is the full light curve. Panel~2 is a periodogram highlighting three peaks used as modes for identifying pulsation phase. Panel~3 depicts the pulsation phase in 5 day windows from the lightcurve of each mode, color-coded by the above highlighted peaks, and converted to a light arrival time delay as per \citet{murphy2014}. An average of these time delays, weighted by the height of the peaks, is overplotted in black. Panel~4 is a Lomb-Scargle periodogram of the time delays to identify periodicities, with colours as per Panel~3. For phase modulation binaries, we expect all modes to exhibit the same pulsation phase modulations, or in the case of two pulsators in a binary, modulations should occur in anti-phase. In the latter case, panel 4 will still show peaks at a single period, corresponding to the binary orbital period, but those peaks have amplitudes corresponding to the binary mass ratio \citep{murphy2016}.}
    \label{fig:shortlist}

\end{figure}

We search for and fit phase modulations from the sample of pulsators in two stages. We first apply the method of \citet{murphy2014} in order to perform a pre-search for pulsational phase modulations from possible binaries. We summarize that approach as follows. First, the frequencies, amplitudes and phases of the modes identified in Sec. \ref{sec:id_pulsators} are measured using the full light curve via a least squares fit of a sum of sine waves. Then, light curves are divided into windows of 5 days, and the phase of each mode in each window is re-measured at the fixed frequencies identified in the previous step. By plotting phase changes as a function of time, we shortlisted all targets by eye with noticeable phase modulations present in more than one mode, and that appear to exhibit the same period and amplitude ($a \sin i$) in each mode (Fig.\,\ref{fig:shortlist}, Panel 3 and 4). We also identify any `PB2' targets where both stars in the binary pulsate; we expect the modes of each star to present phase modulations at the same orbital period but in anti-phase (see \citealt{murphy2014}). We also separate out systems for which the orbital period appears to be longer than the available baseline of \textit{TESS} photometry, or those for which the available data appear to be insufficient to uniquely constrain the orbital parameters.

After the pre-search, we identify systems whose orbits can be fitted with the available data. We forward model their pulsation timing variations using the modes of the light curve identified during the pre-search, based on the algorithm used in {\tt maelstrom} \citep{hey2020}. We implement our phase modulation forward model in \texttt{jax}, solving Kepler's equations with the package \texttt{jaxoplanet} \citep{jaxoplanet}. Our model uses six free parameters: $P_{\rm orb}$, the  period; $a_1 \sin(i)/ c$, the light travel time (where $a_1 \sin(i)$ is the projected semi-major axis); $\omega$, the argument of periapsis\footnote{\citet{murphy2015} and papers in that series refer to this quantity as $\varpi$ so as not to cause confusion with angular oscillation frequencies; we do not use the latter here.}; $e$, the eccentricity; $\phi$, the phase of the periapsis; and $\nu$, a vector parameterizing the frequency of each mode. We first individually optimize the light travel time, $a_1 \sin(i)/ c$ and the orbital period for each mode to confirm that the different modes have similar optimization values. For true phase modulation binaries, we expect all modes to be consistent with the same value of $a_1 \sin(i)/ c$ due to the common motion of the pulsator. This excludes false positives due to aliasing of two very nearby modes, which can manifest as spurious phase modulations in one of the modes at the ``beat" frequency of the two modes. An exception to this are `PB2' systems where both stars in the binary exhibit pulsation modes. From each mode's individual time delay, we evaluate whether different modes are consistent with opposite $a \sin(i) / c$, which are indicative of `PB2' systems. We separate and refit the PB2 systems with a model that pins each mode to either $a_1 \sin(1) / c$ or $a_2 \sin(1) / c$ and then jointly fits the system's orbital parameters according to the method of \citet{hey2020}. 

For each parameter, we use wide priors centered at the optimization value, except for $\nu$, for which we use a normal prior centered on our Lomb--Scargle peaks with a standard deviation of 0.01\,d$^{-1}$. For period and $a \sin i$, we use a lognormal prior with a wide standard deviation. For the other parameters, we use uniform priors through the entire physical range of parameter space. We sample our model with the No U-Turn Sampler \citep{Hoffman2011} as implemented in the package \texttt{numpyro}.

We obtain medians and 1 $\sigma$ intervals from the posterior distributions of each parameter from our MCMC fit. Some binaries have a period longer than the available photometric baseline. We nevertheless attempt to fit a majority of these systems with our methodology as well, with the caveat that the period, eccentricity, and $a \sin i$ may be degenerate and/or unreliable (see discussion in \citealt{murphy2018}). Although the parameters for these systems should be approached with caution, they may be amenable to follow-up either from the ground or space, including with more \textit{TESS} data in the future.

\subsection{Period recovery in undersampled systems}
\label{sec:ir}
\begin{figure}
\label{sectors}
\includegraphics[width = \hsize]{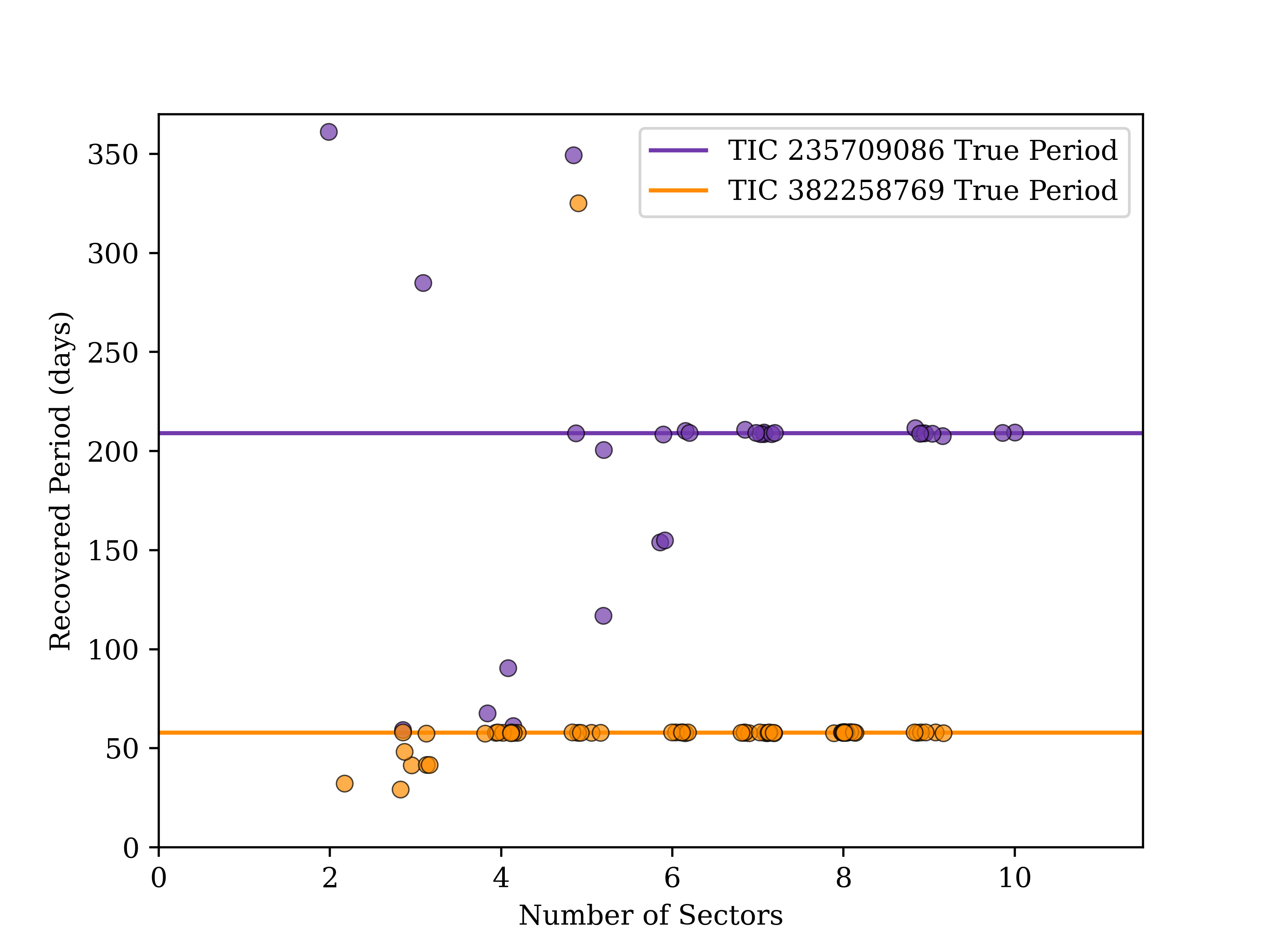}
\caption{Recovered period compared to the expected period as a function of number of sectors of \textit{TESS} data used in the fit. The purple and orange line and points represent the data for TIC 382258769 and TIC 235709086 respectively. The line marks the period recovered with the full set of \textit{TESS} data, and each point represents the period recovered by one subsample of the full data described in Section \ref{sec:ir}. The points are jostled slightly on the x-axis for distinguishability. For seven sectors or more of \textit{TESS} data used, we consistently and reliably estimate the period for both cases tested.}
\label{fig:sectortest}
\end{figure}

In order to identify the minimum number of \textit{TESS} sectors required to identify binaries, we perform a recovery exercise on two binaries with well-defined orbital parameters and complete photometric coverage over the available baseline: TIC~382258769 and TIC~235709086. These two cases were chosen to represent a short period (58\,d, with 330-d baseline) and a longer period (209\,d, 360-day baseline) case. We simulate light curves with reduced duty cycle by randomly excluding sectors from the middle of the light curve, while keeping the total observation baseline constant (we always keep the first and last sectors). For each one of these reduced duty cycle light curves, we perform the MCMC fit described in Section \ref{sec:phasemod} and attempt to recover the orbital parameters of the binaries. Figure \ref{fig:sectortest} shows the recovered period against the correct period as the number of sectors used is varied. For the shorter period binary, we are able to recover the correct period with only 6 sectors of observation, while the longer period binary needs at least 7 sectors of observation to correctly measure the period. Given the relatively low occurrence rate of short period binaries (50--200\,d) we expect negligible gain in detection yields by including stars observed by \textit{TESS} for only 6 sectors. We therefore set 7 sectors as our minimum requirement to search for phase modulations. 
%We identify 7 sectors as a minimum required to robustly identify the correct period. Below this, the true period value lies outside the error bars of the recovered value in some cases for both binaries tested. Thus, with fewer than 7 sectors we cannot reliably recover the correct orbit for either binary. We therefore set 7 sectors as our minimum requirement to search for phase modulations.

\section{Binary Sample}
\label{sec:binary_sample}
\begin{figure}
\centering
\includegraphics[width=0.95\linewidth]{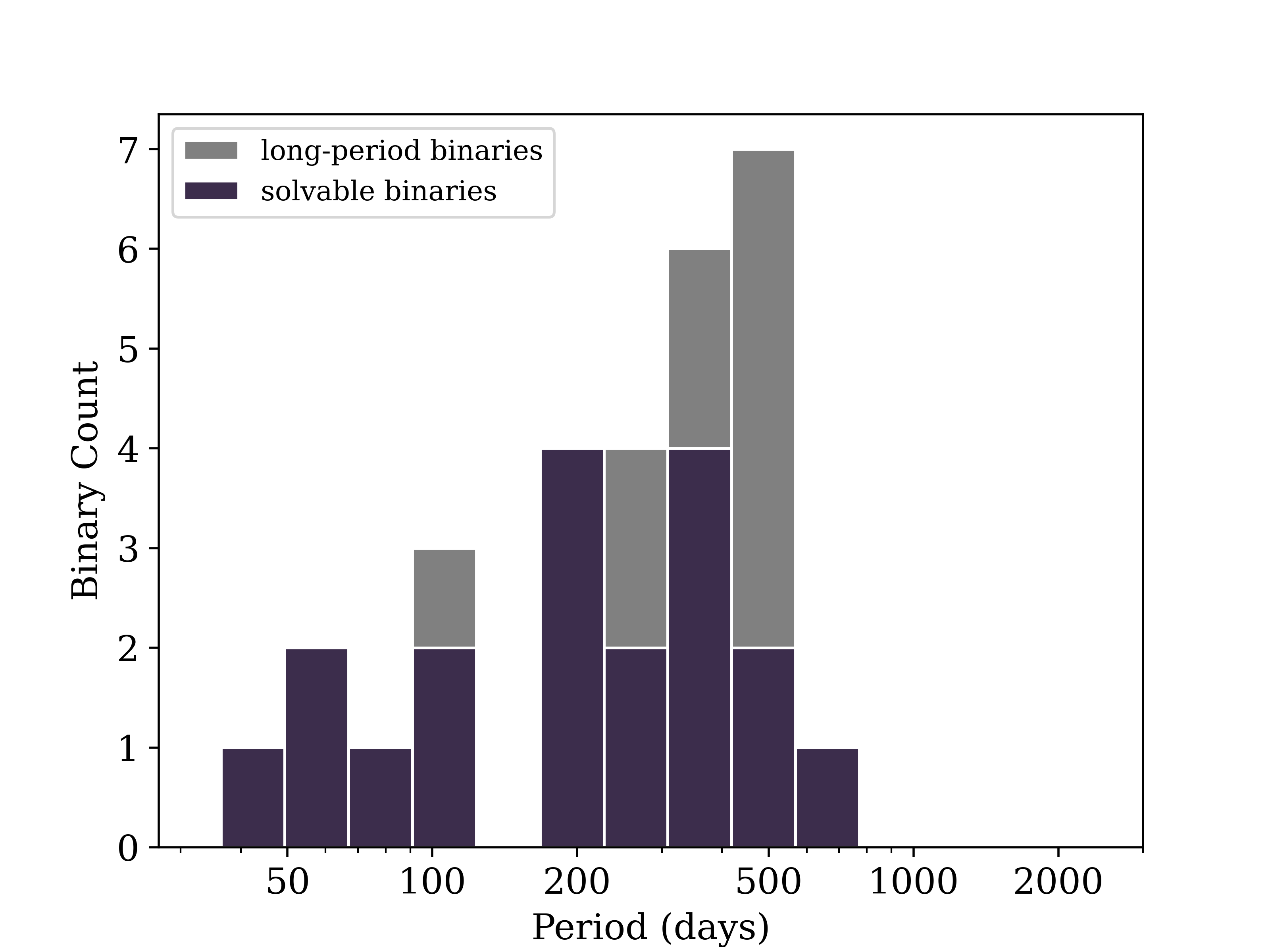}
\includegraphics[width=0.95\linewidth]{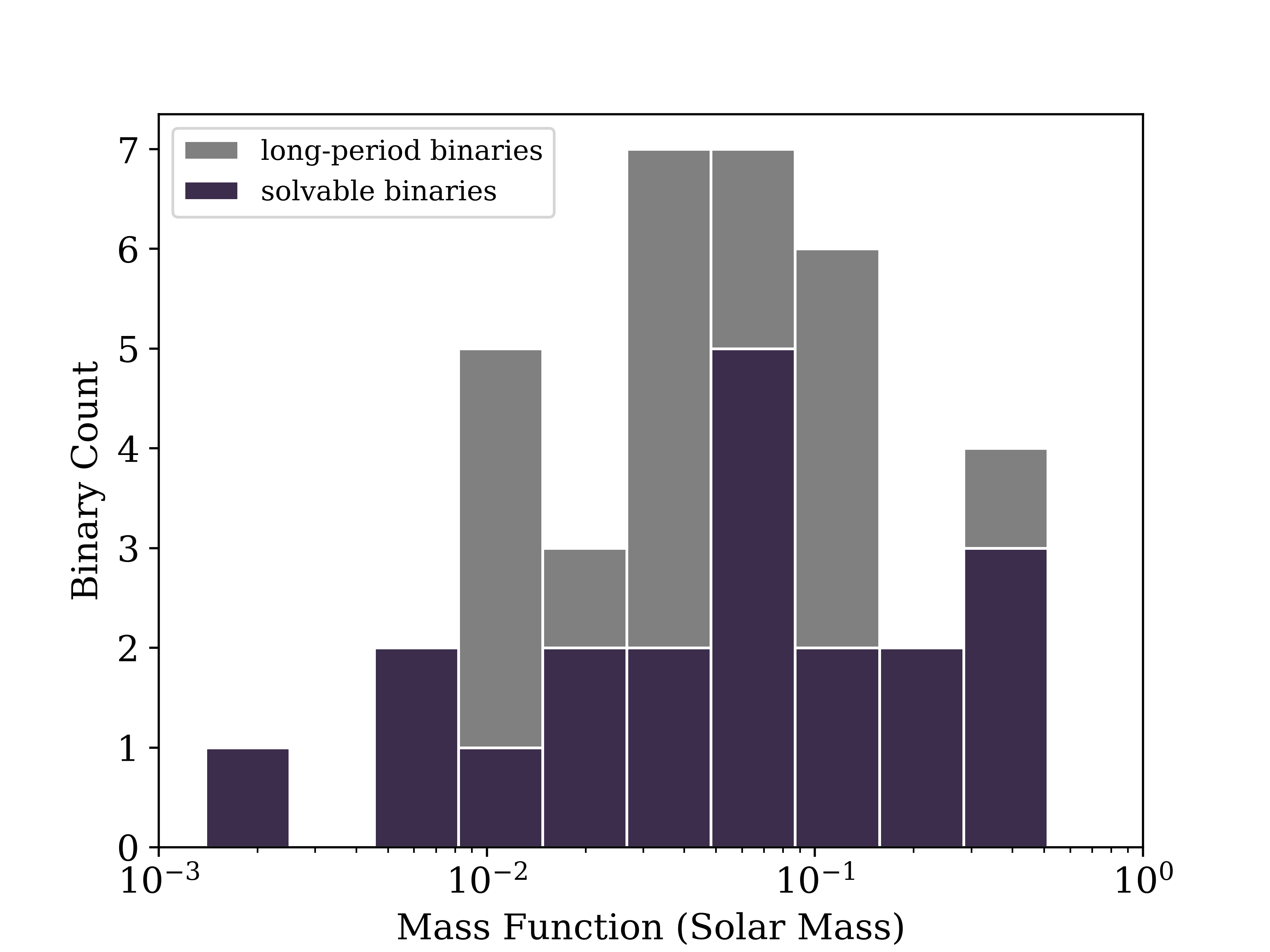}
\includegraphics[width=0.95\linewidth]{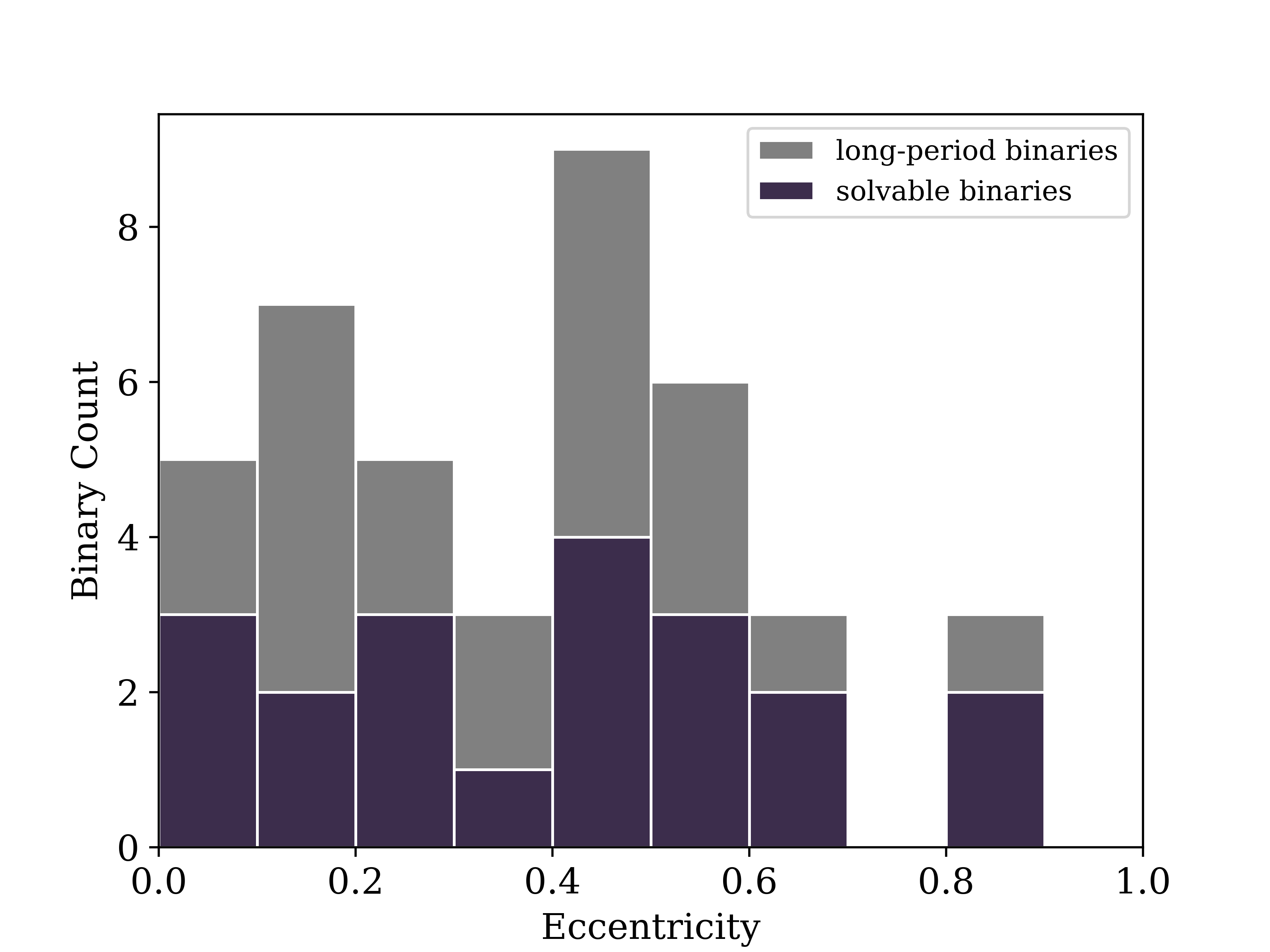}
\caption{Histograms of parameters from the MCMC fits of the phase modulation binary sample. The top, middle, and bottom plots show the period, mass function and eccentricity, respectively. Solvable binaries are plotted in purple and long-period binaries are in gray. We note that the orbital parameters for these long-period binaries may be degenerate due to the lack of phase coverage for these targets.}
\label{fig:hists}
\end{figure}

\begin{figure*}
\centering
\includegraphics[width=0.3\textwidth]{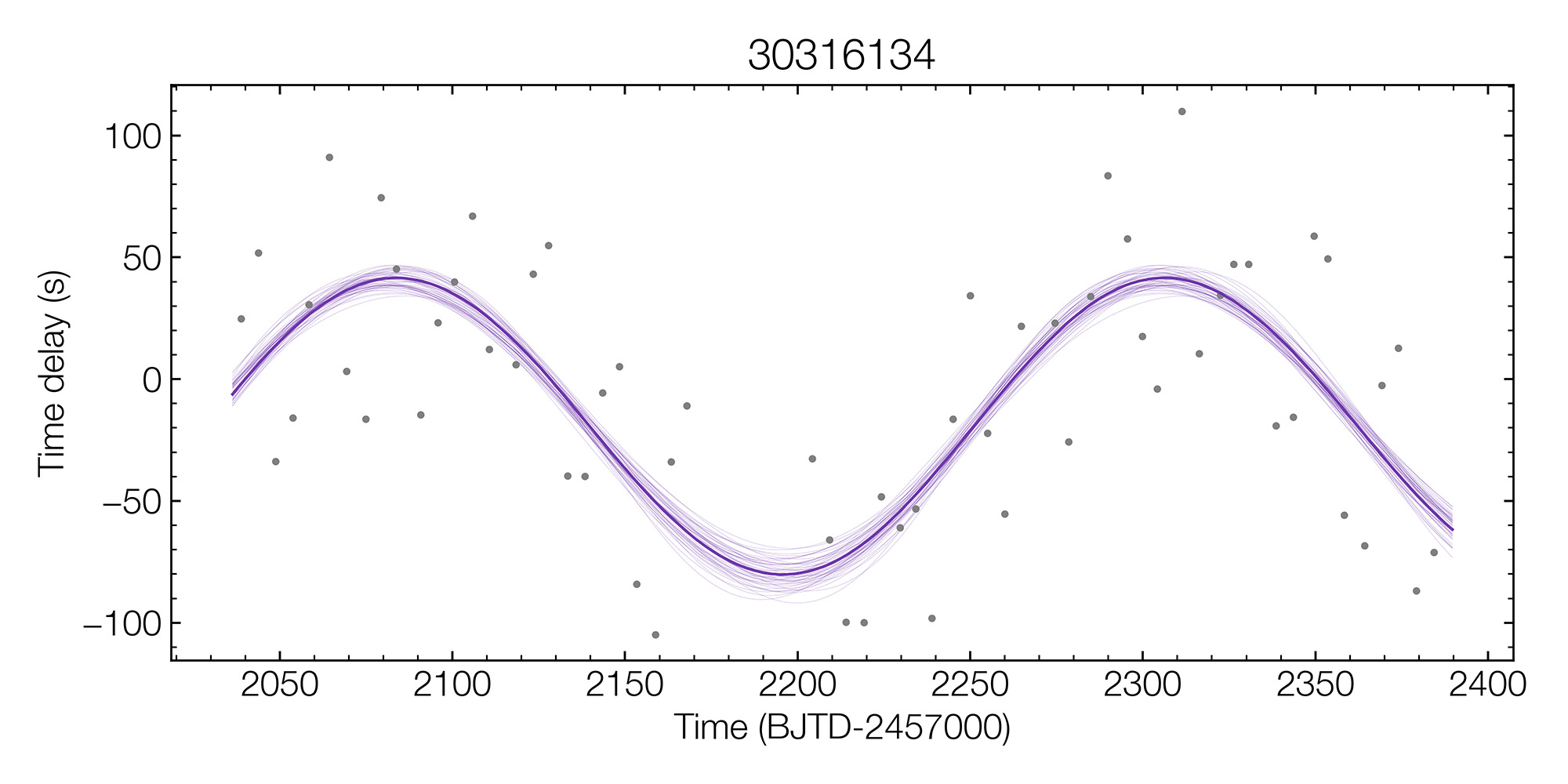}
\includegraphics[width=0.3\textwidth]{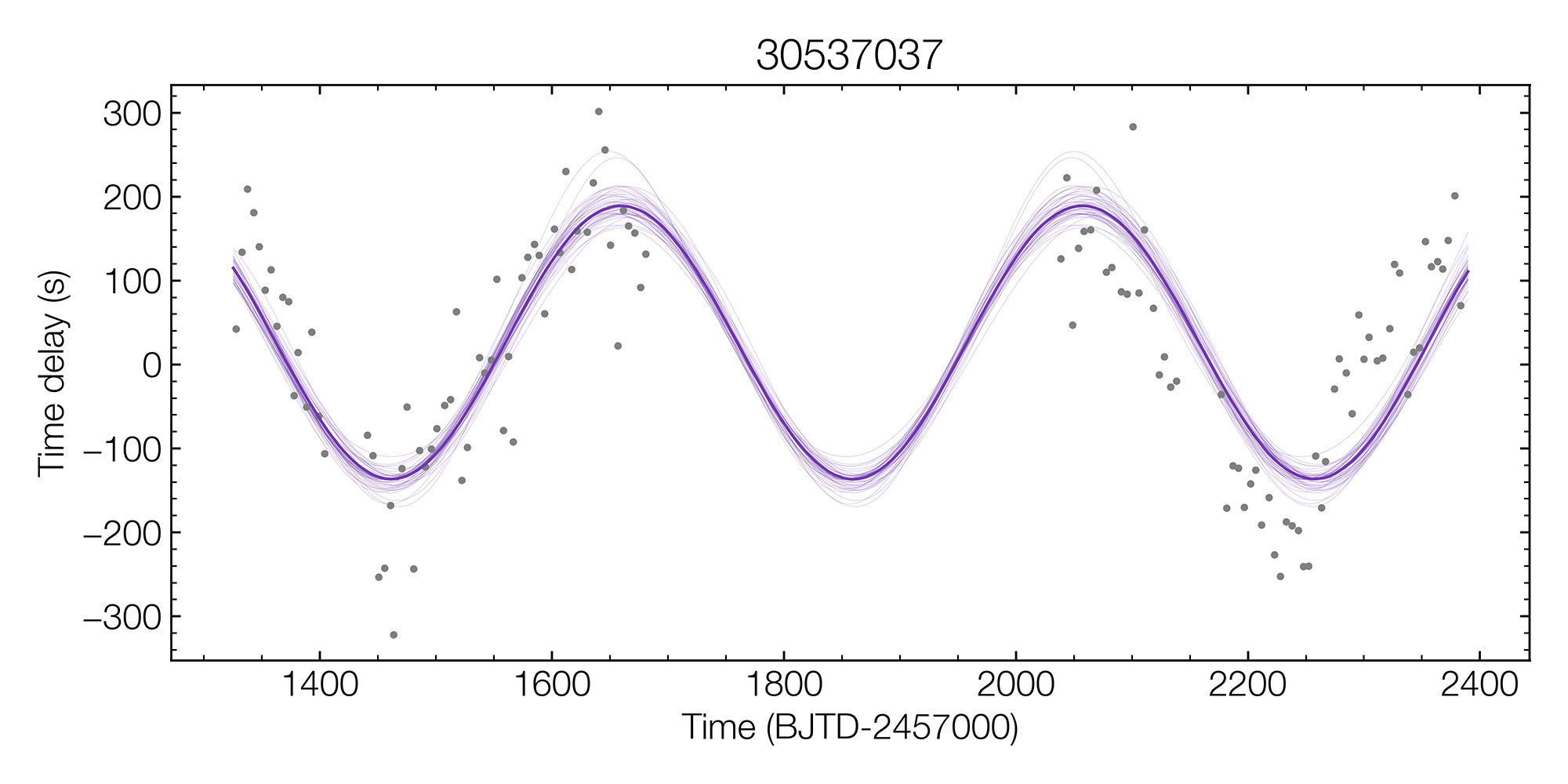}
\includegraphics[width=0.3\textwidth]{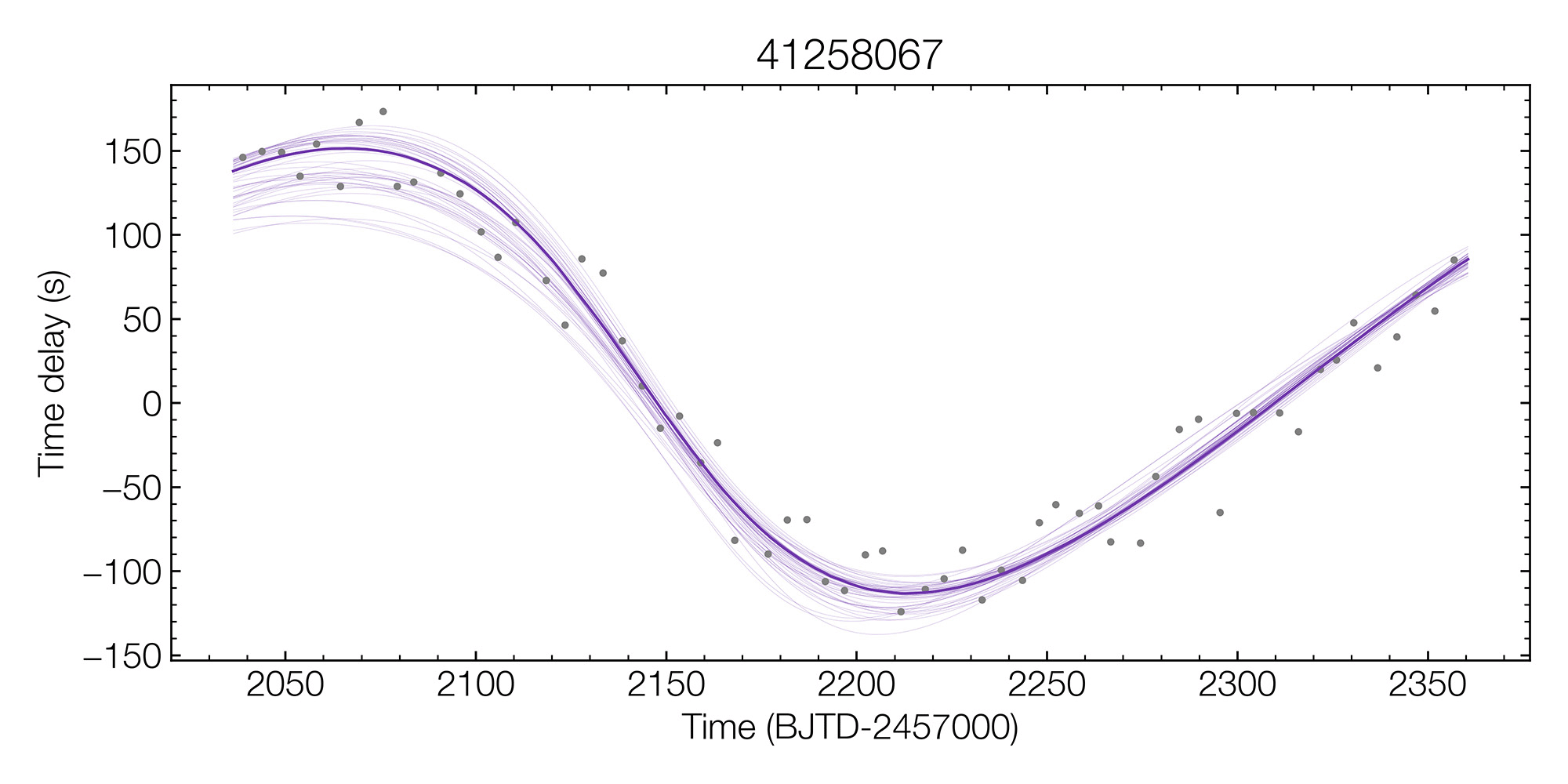}
\includegraphics[width=0.3\textwidth]{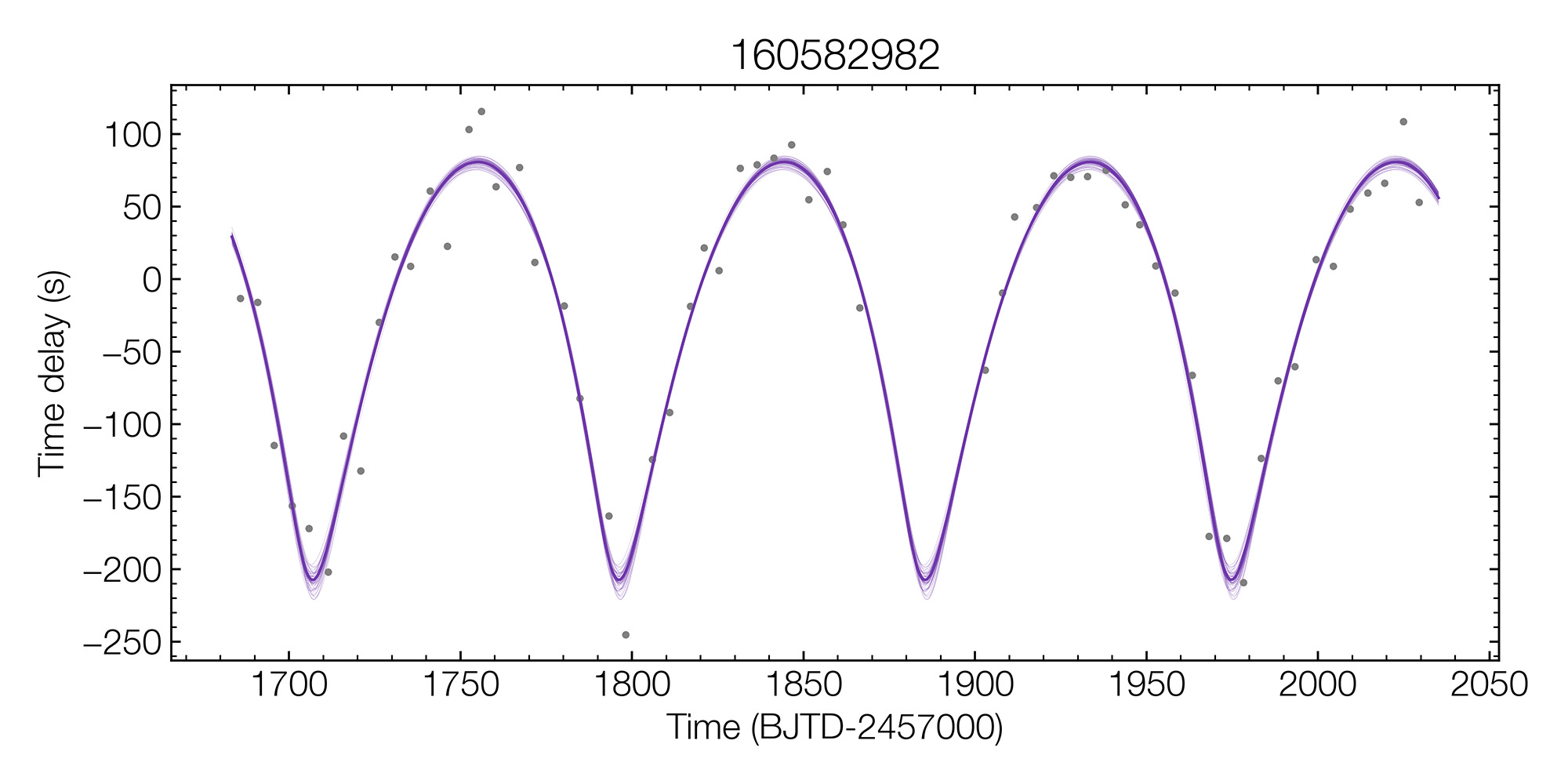}
\includegraphics[width=0.3\textwidth]{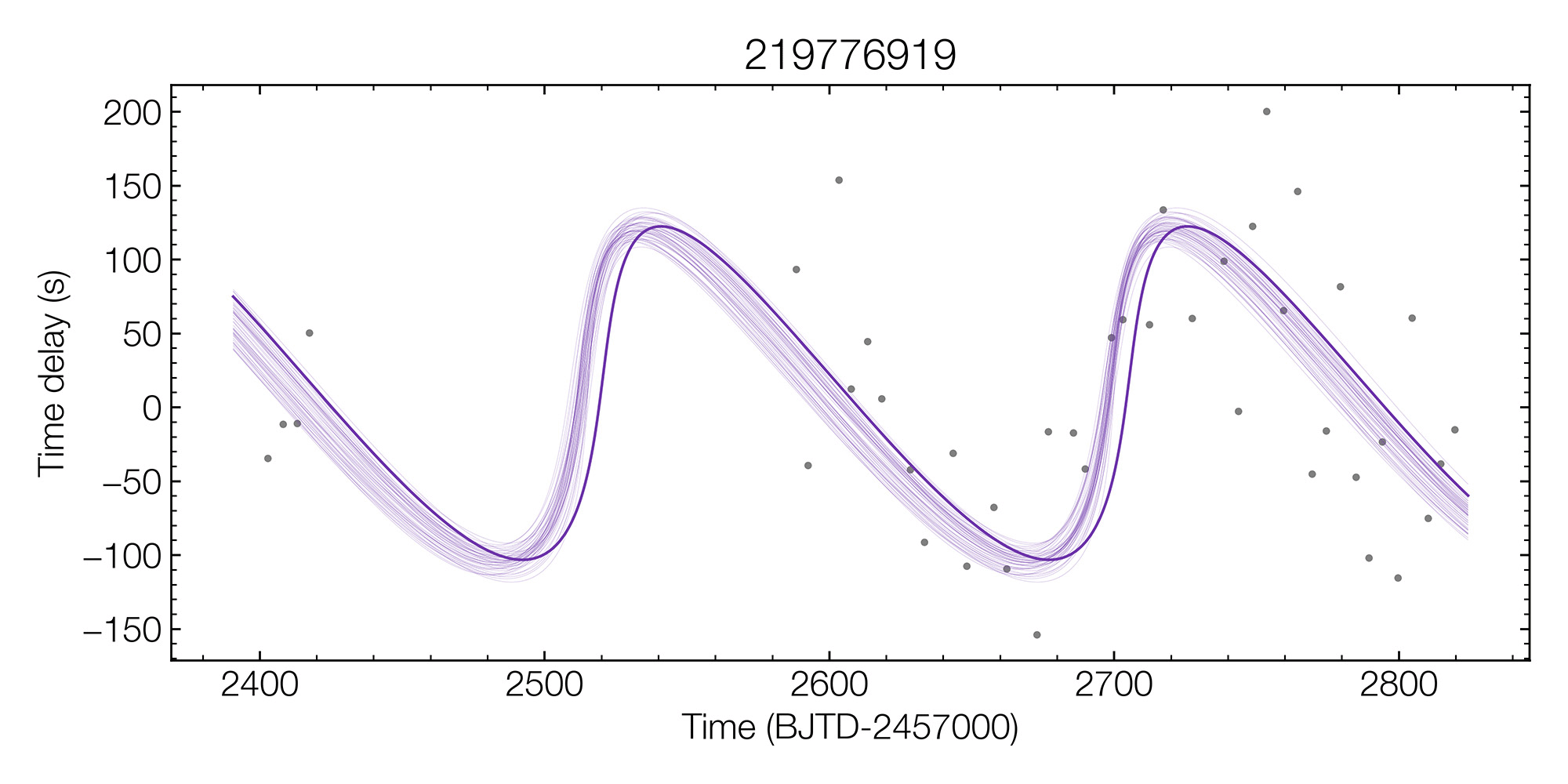}
\includegraphics[width=0.3\textwidth]{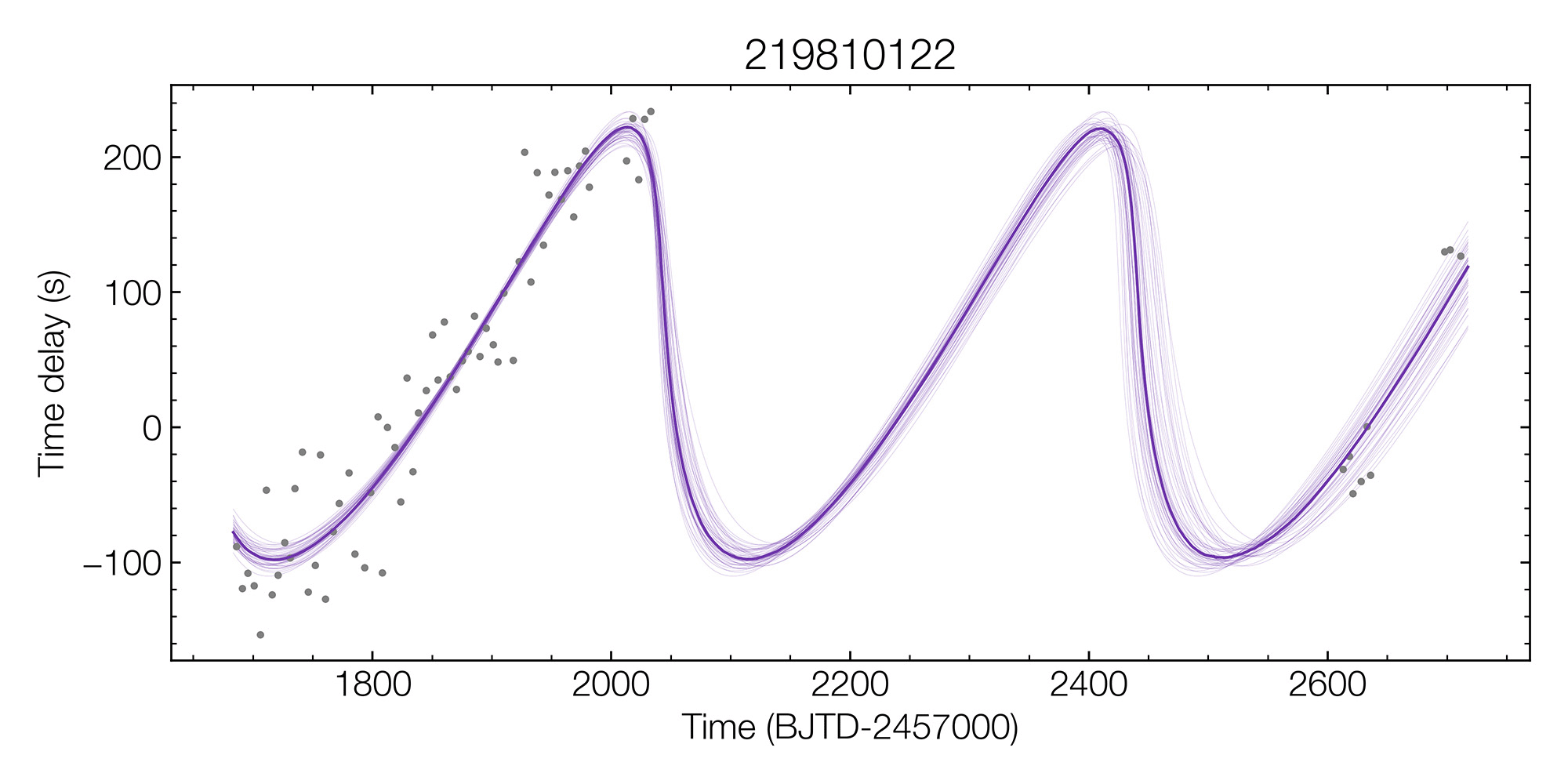}
\includegraphics[width=0.3\textwidth]{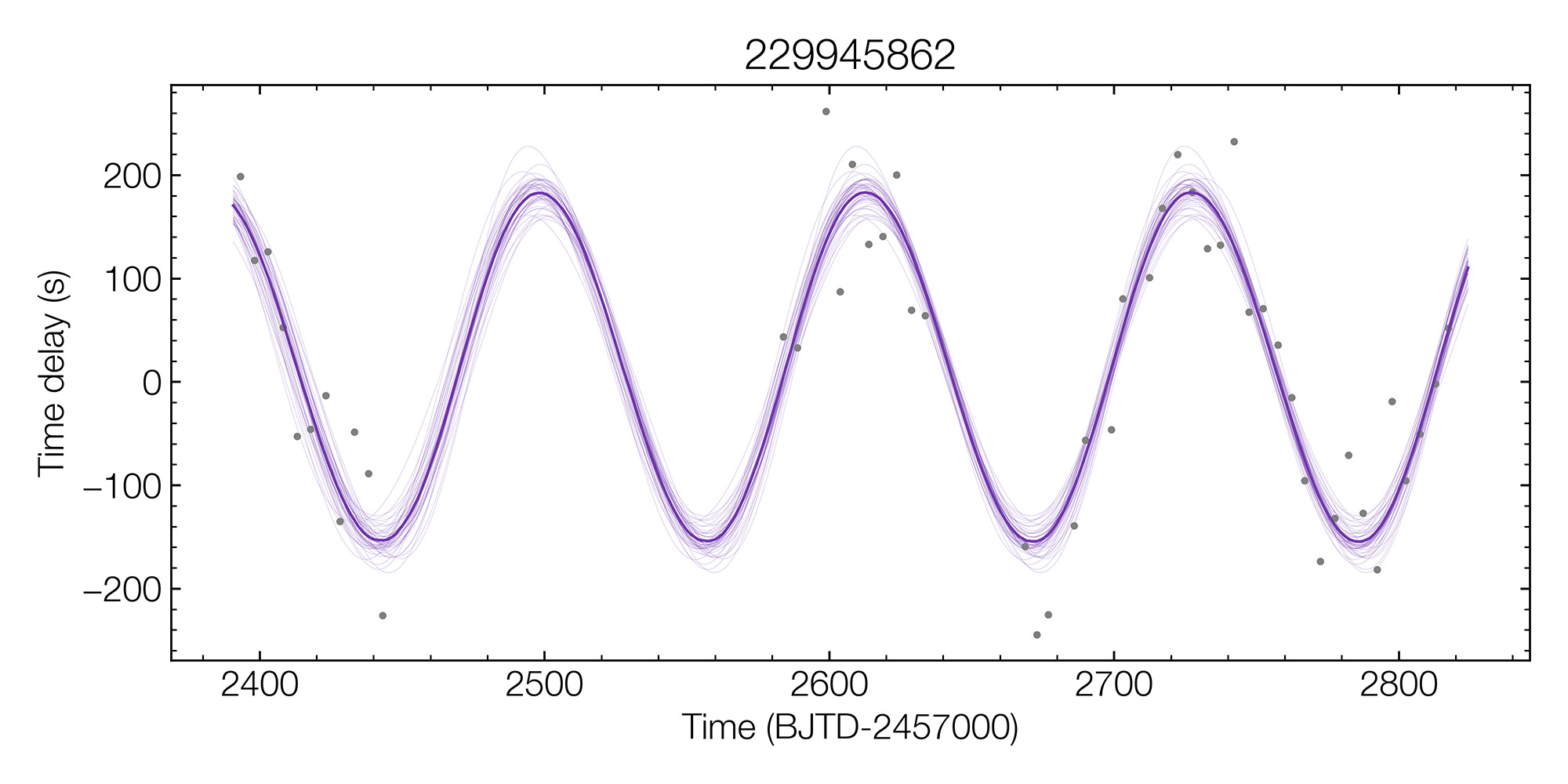}
\includegraphics[width=0.3\textwidth]{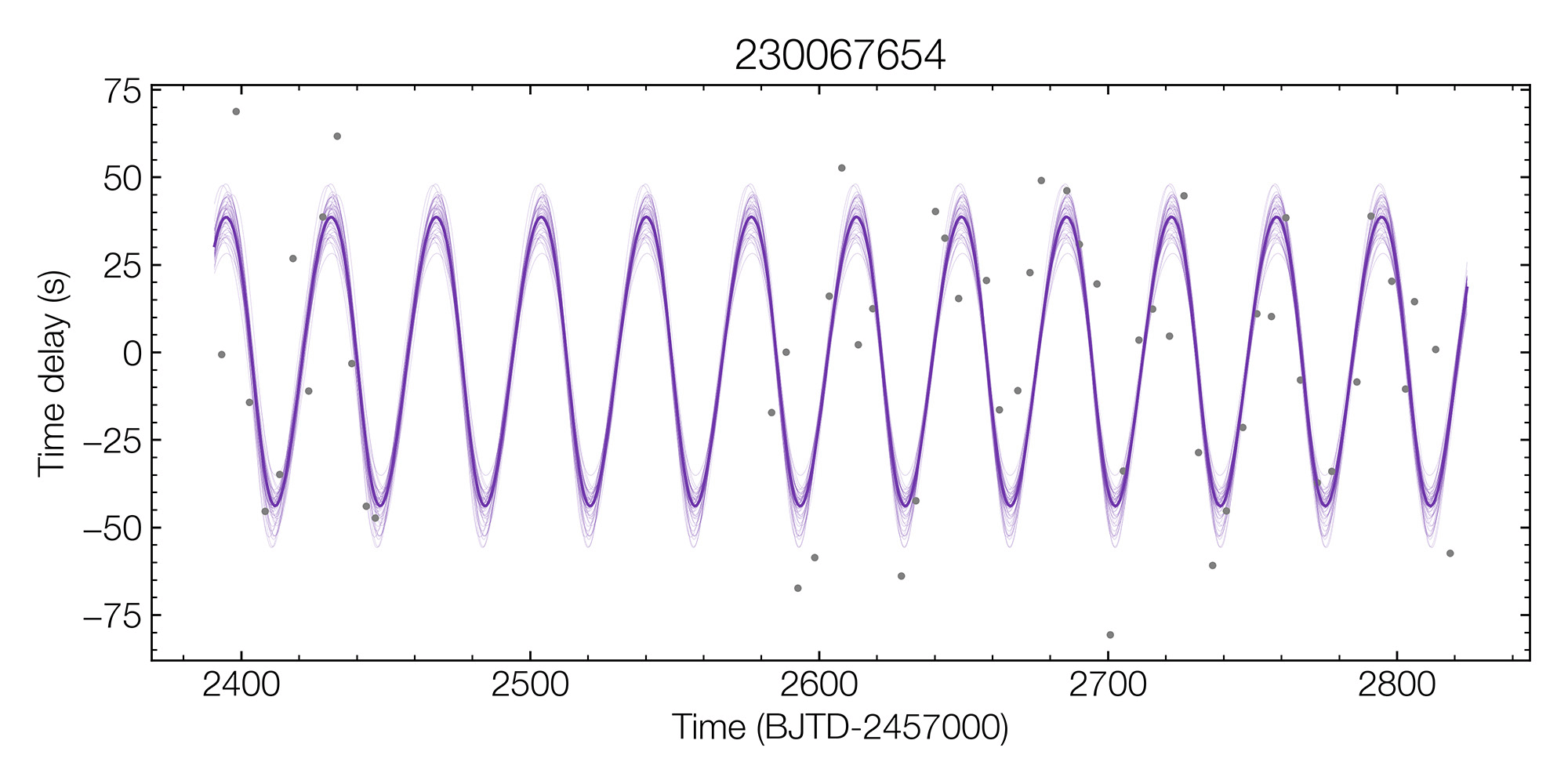}
\includegraphics[width=0.3\textwidth]{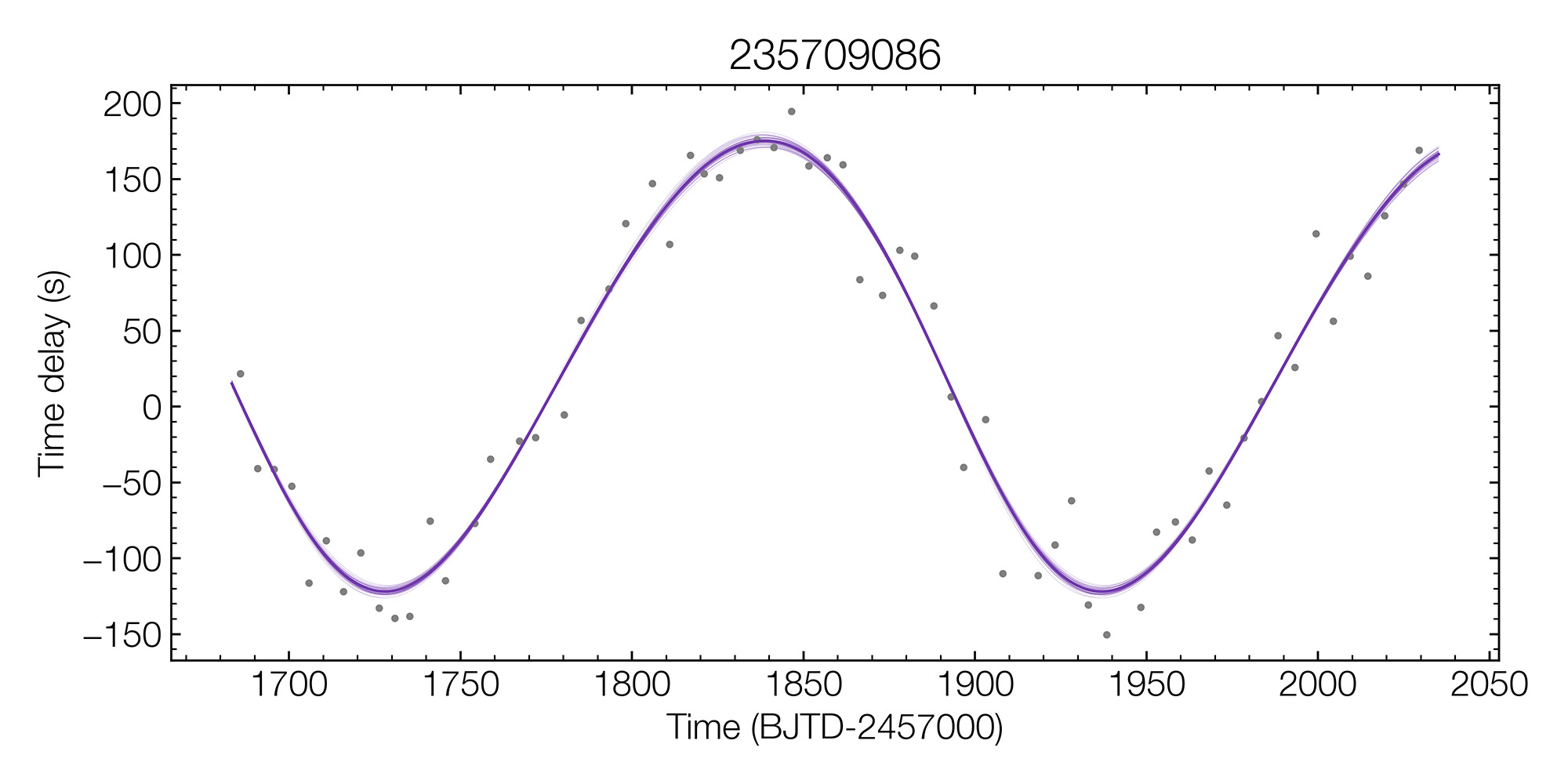}
\includegraphics[width=0.3\textwidth]{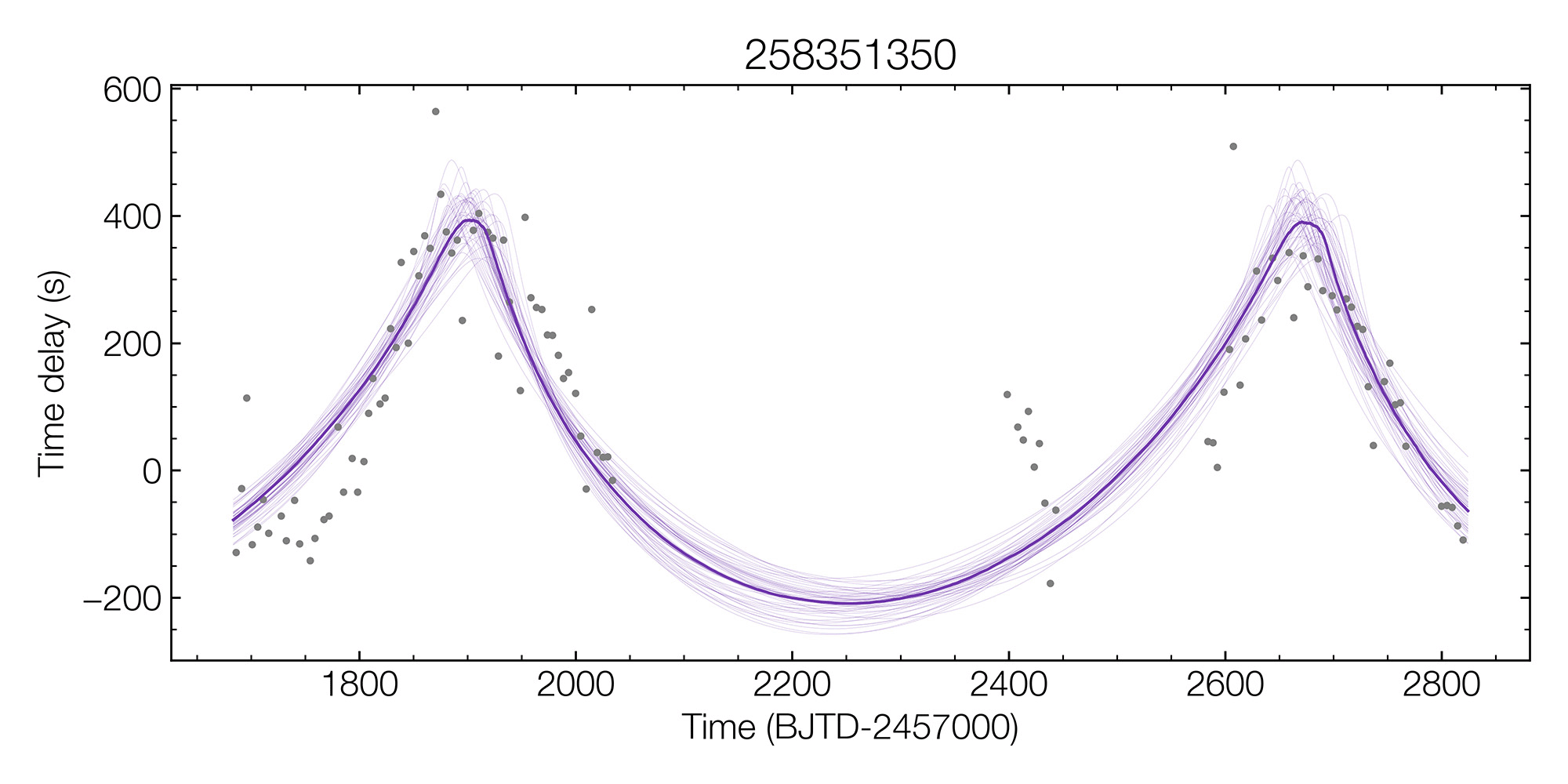}
\includegraphics[width=0.3\textwidth]{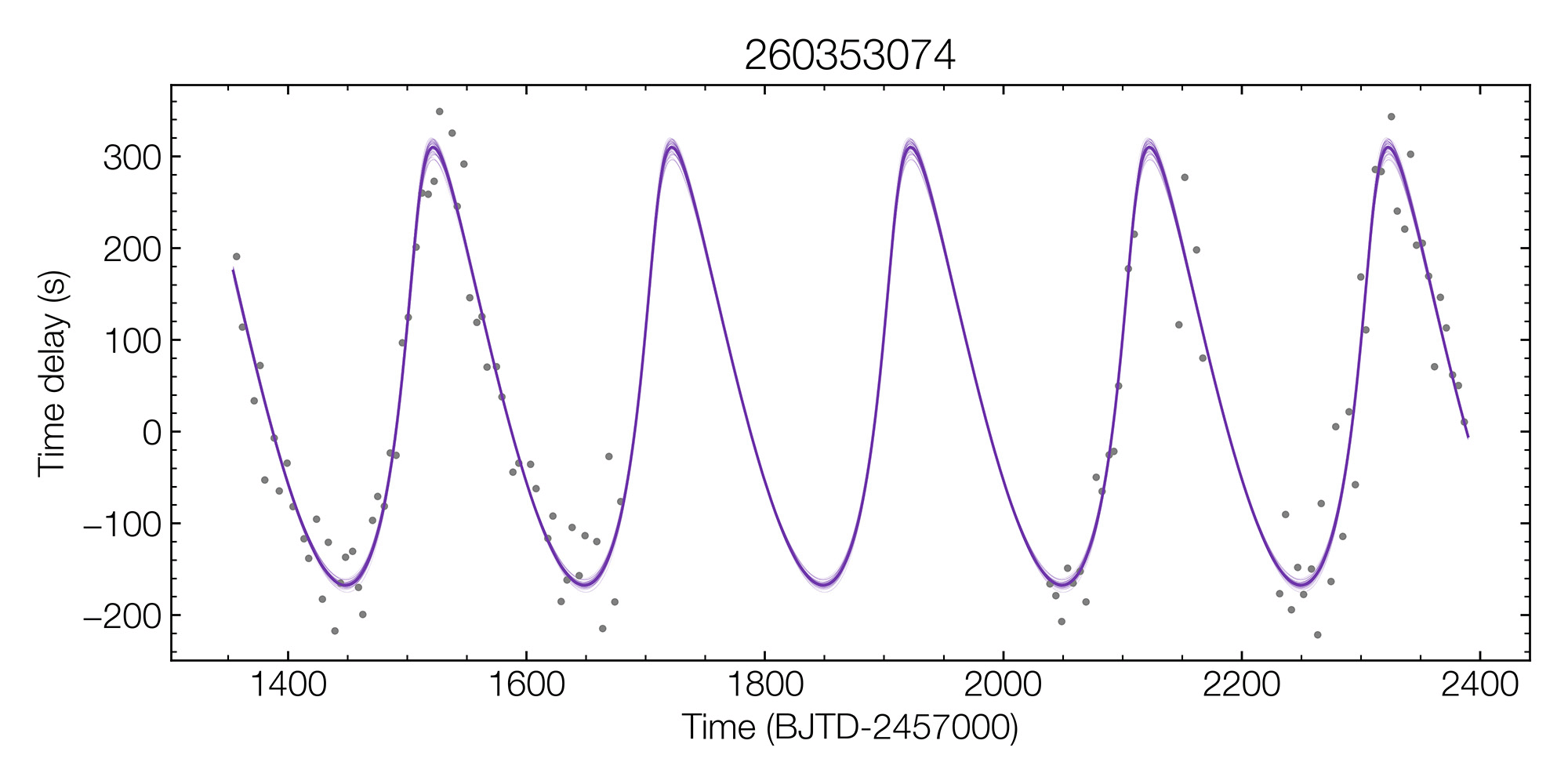}
\includegraphics[width=0.3\textwidth]{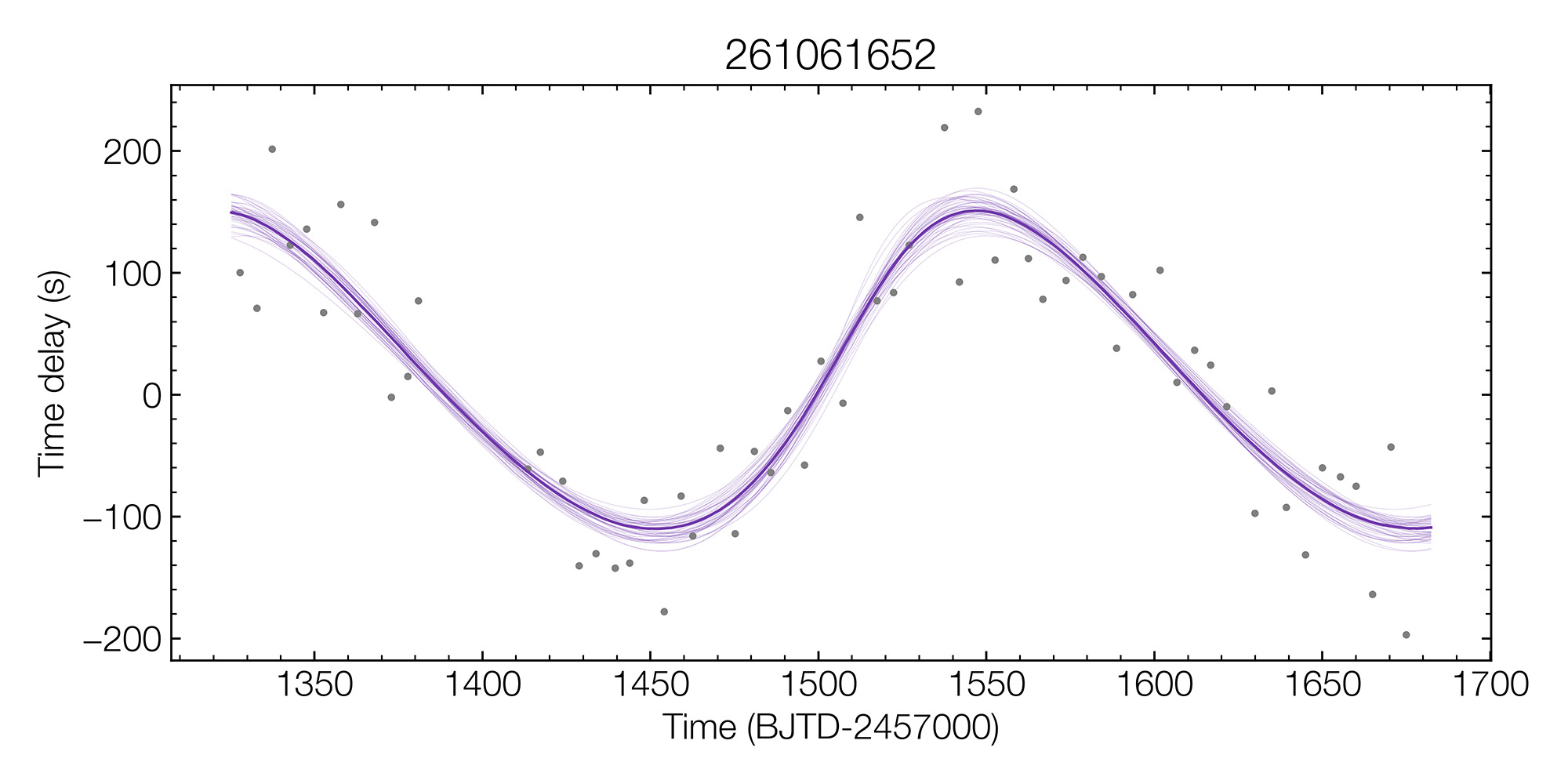}
\includegraphics[width=0.3\textwidth]{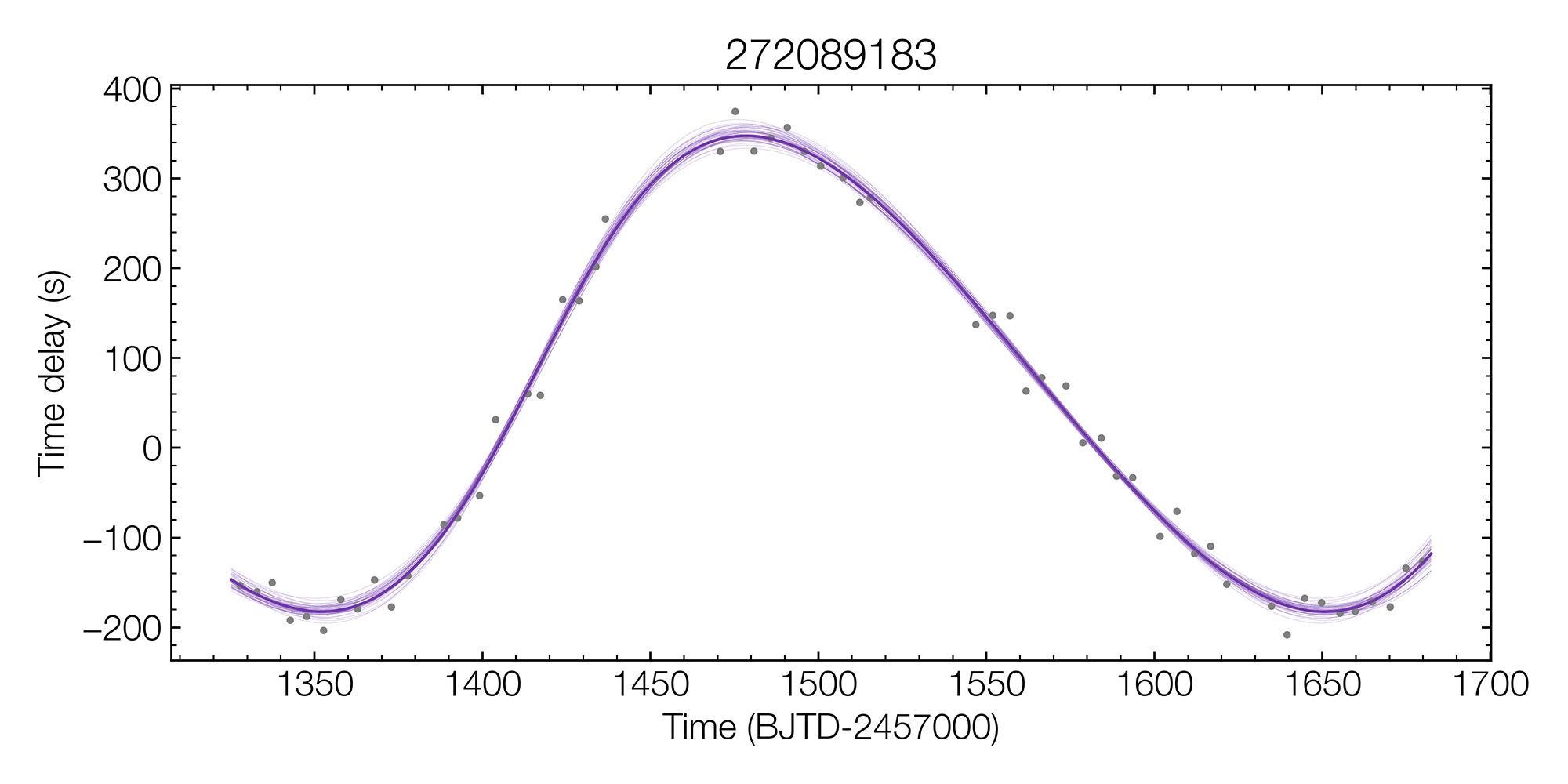}
\includegraphics[width=0.3\textwidth]{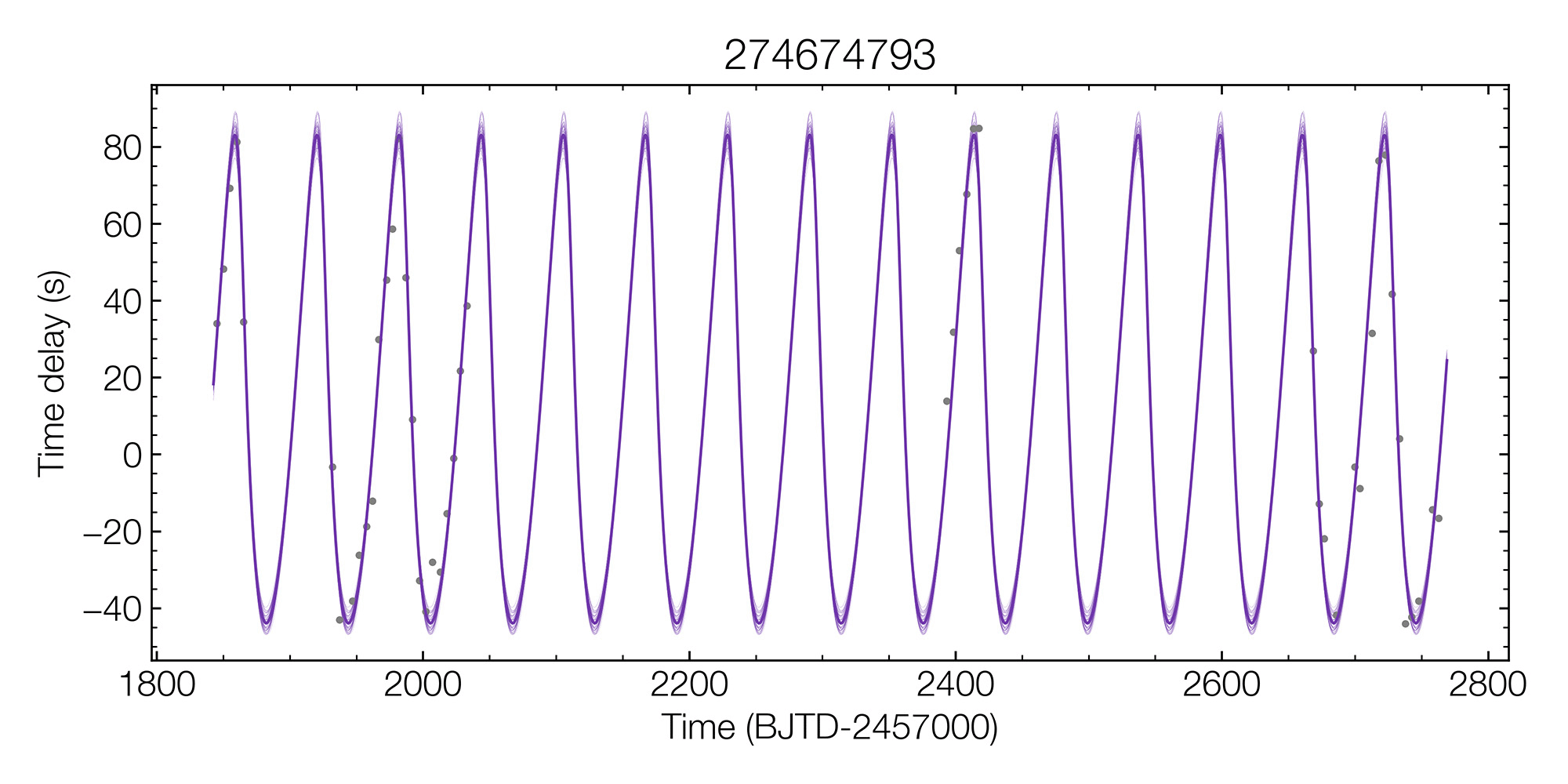}
\includegraphics[width=0.3\textwidth]{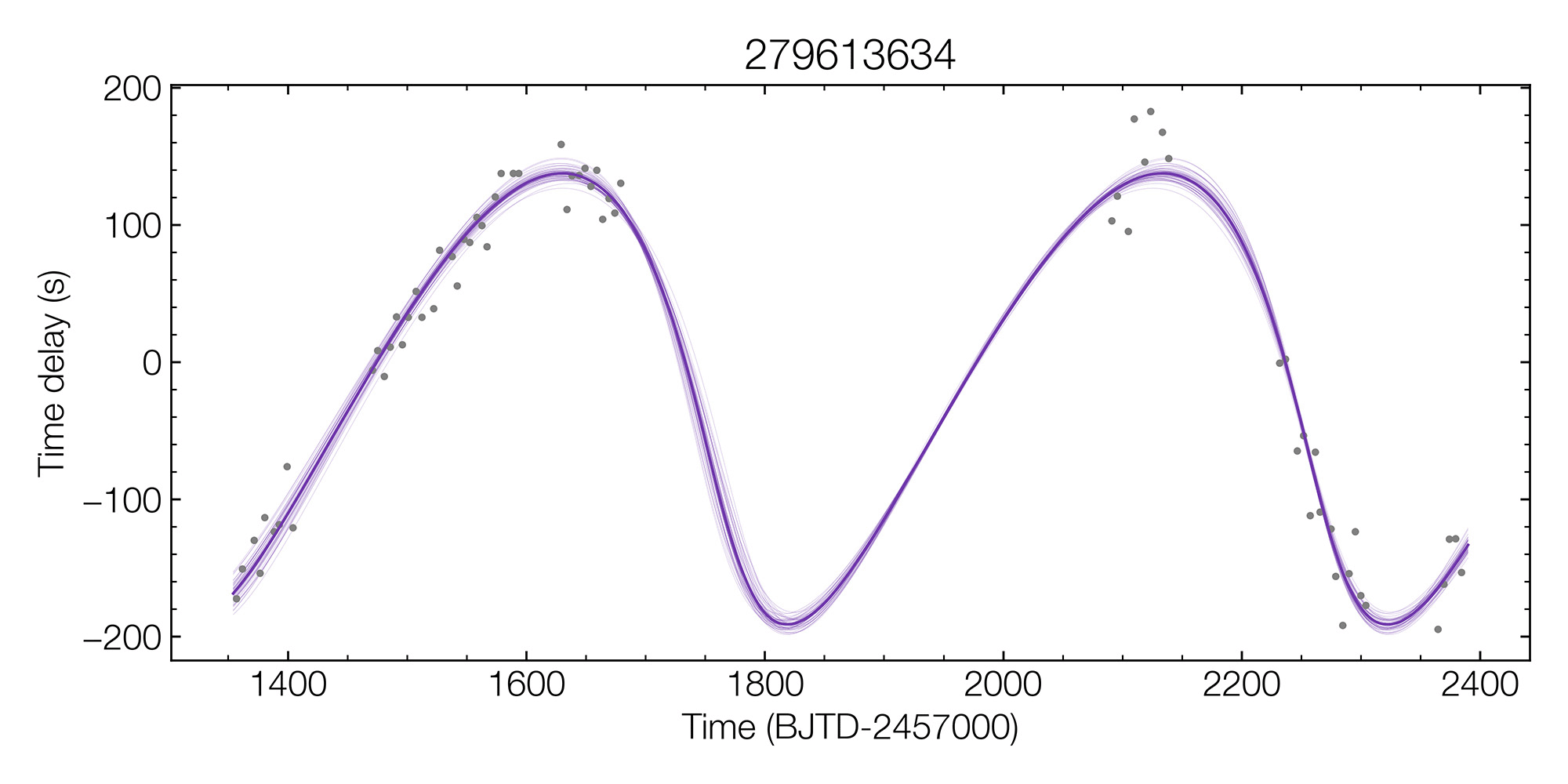}
\includegraphics[width=0.3\textwidth]{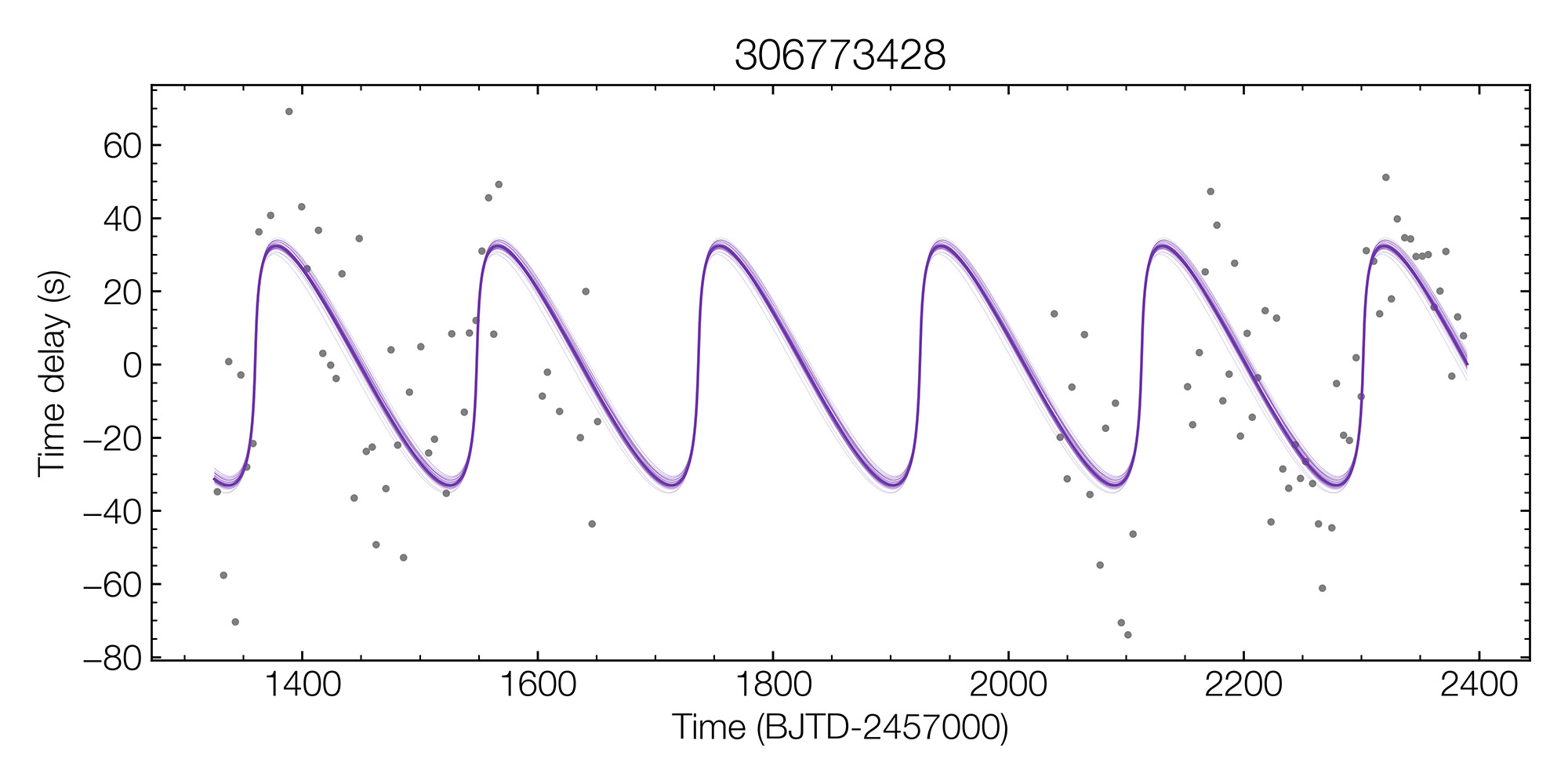}
\includegraphics[width=0.3\textwidth]{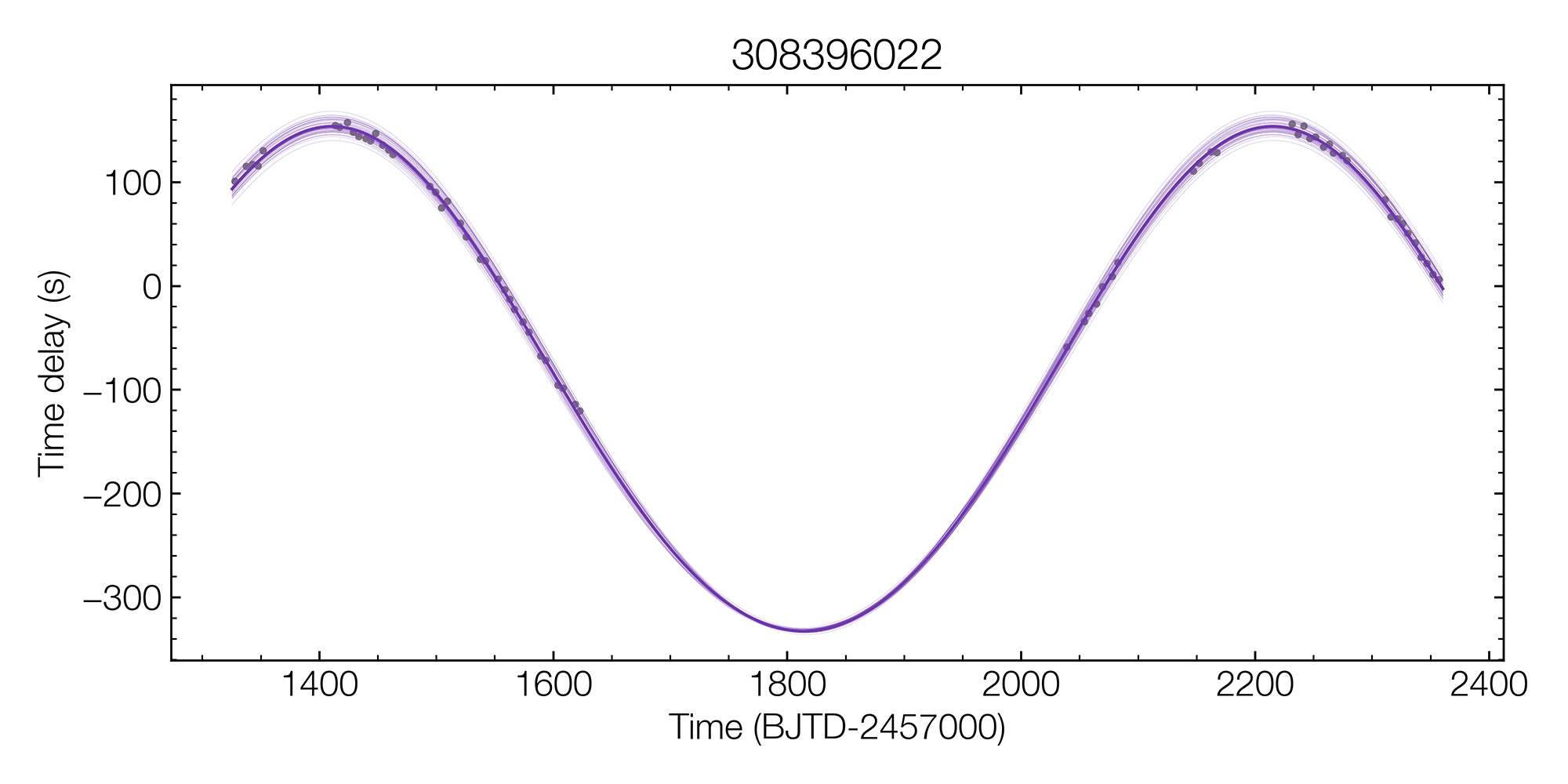}
\includegraphics[width=0.3\textwidth]{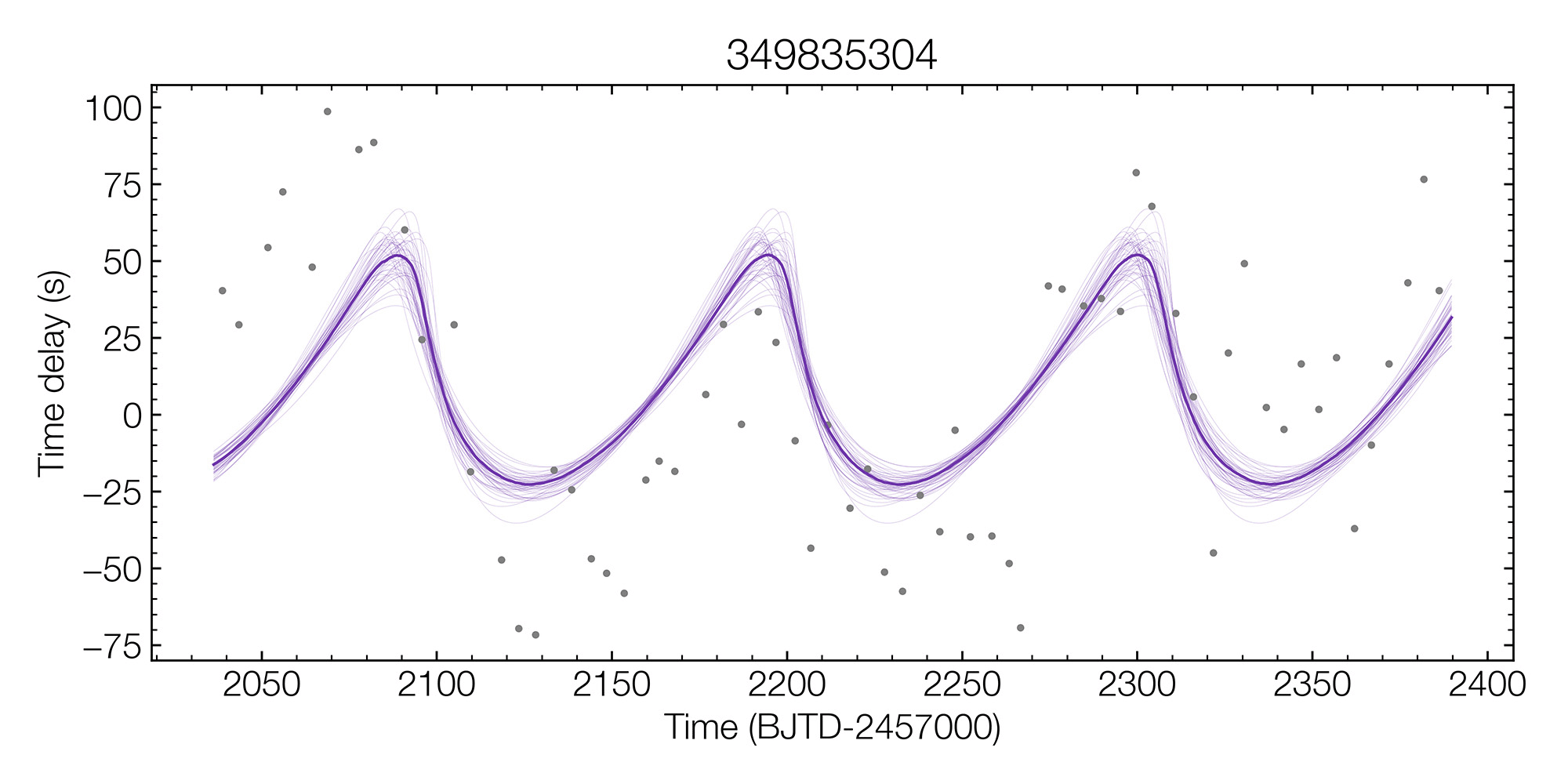}
\includegraphics[width=0.3\textwidth]{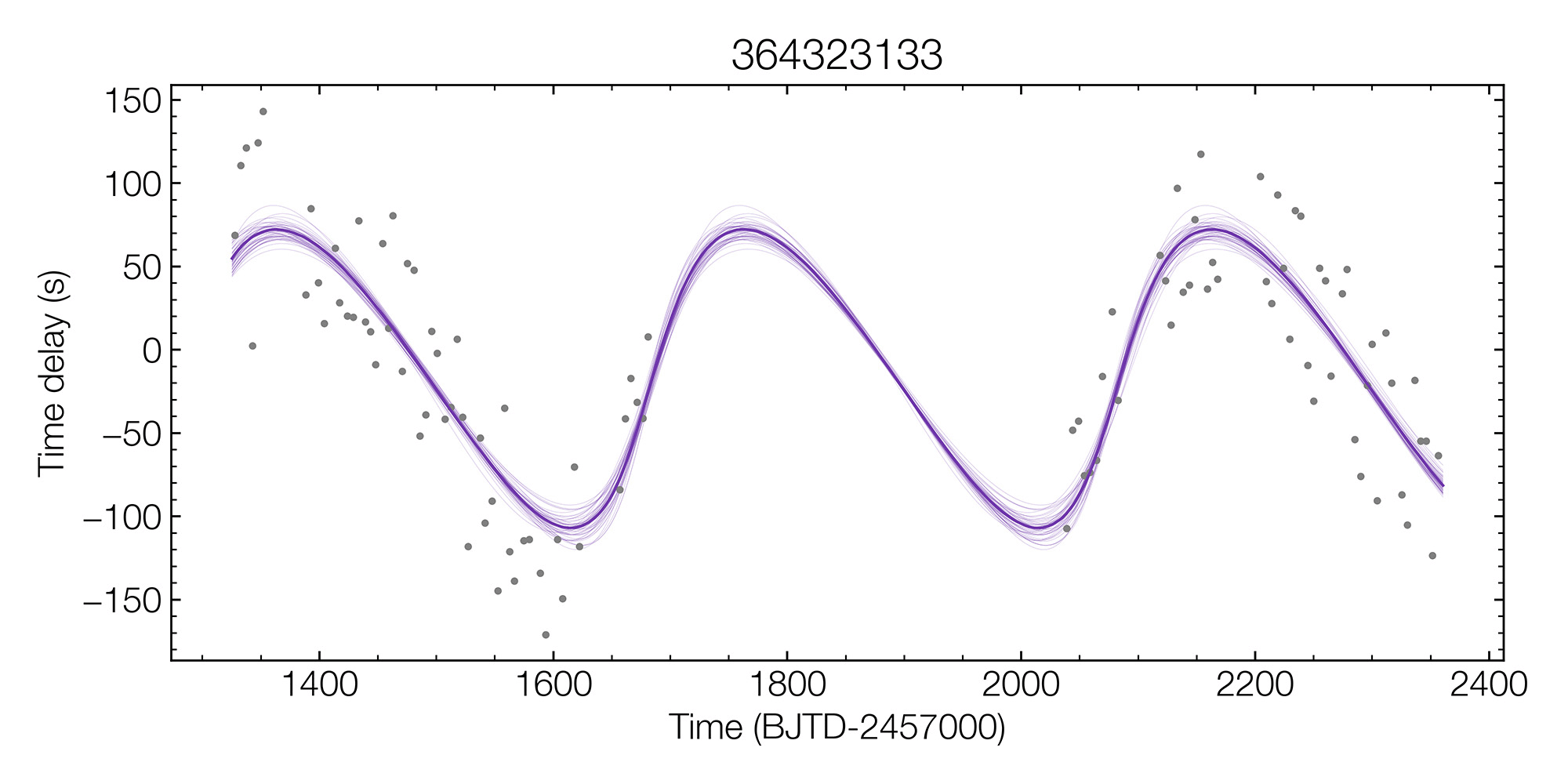}
\includegraphics[width=0.3\textwidth]{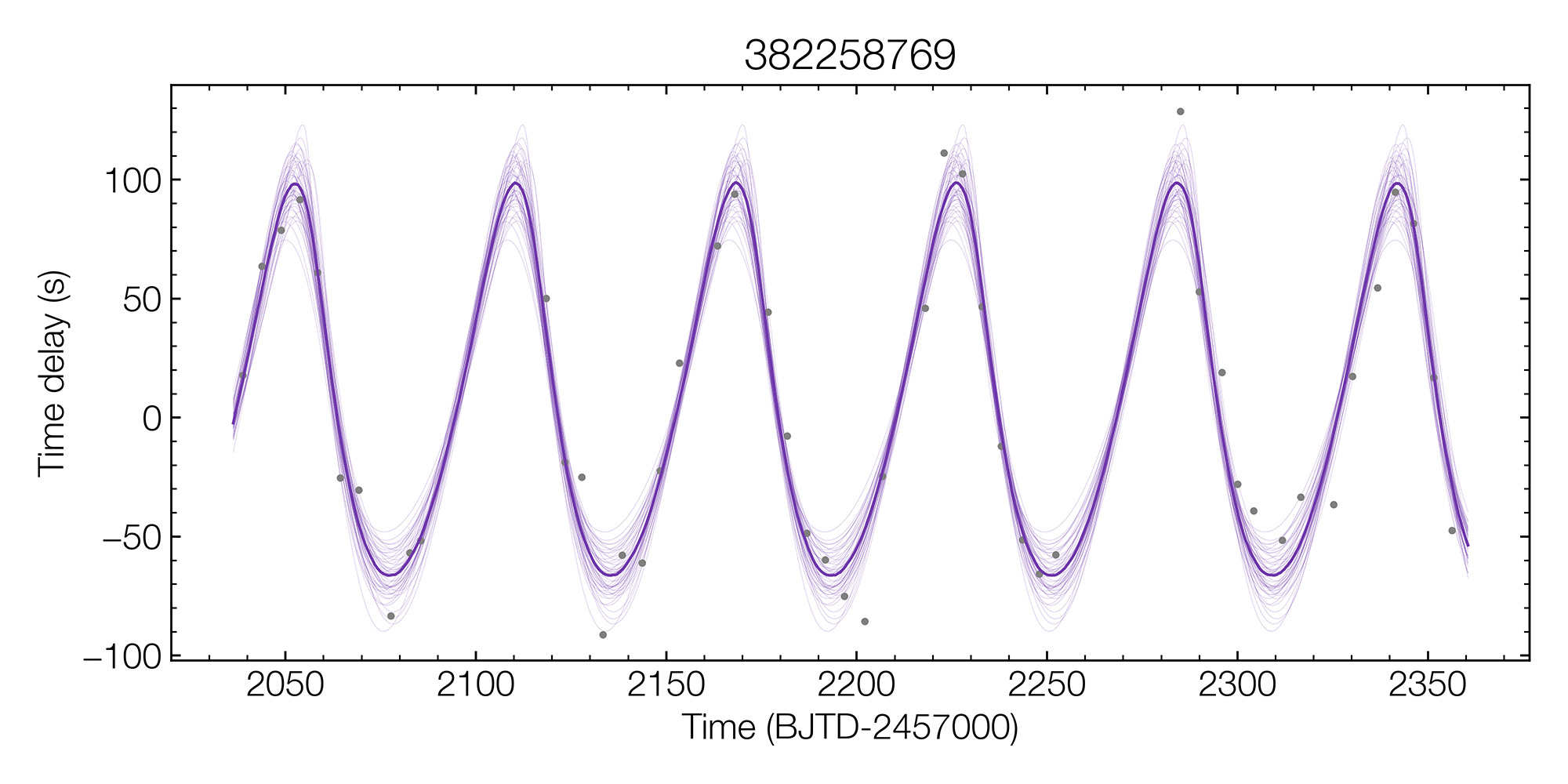}
\includegraphics[width=0.3\textwidth]{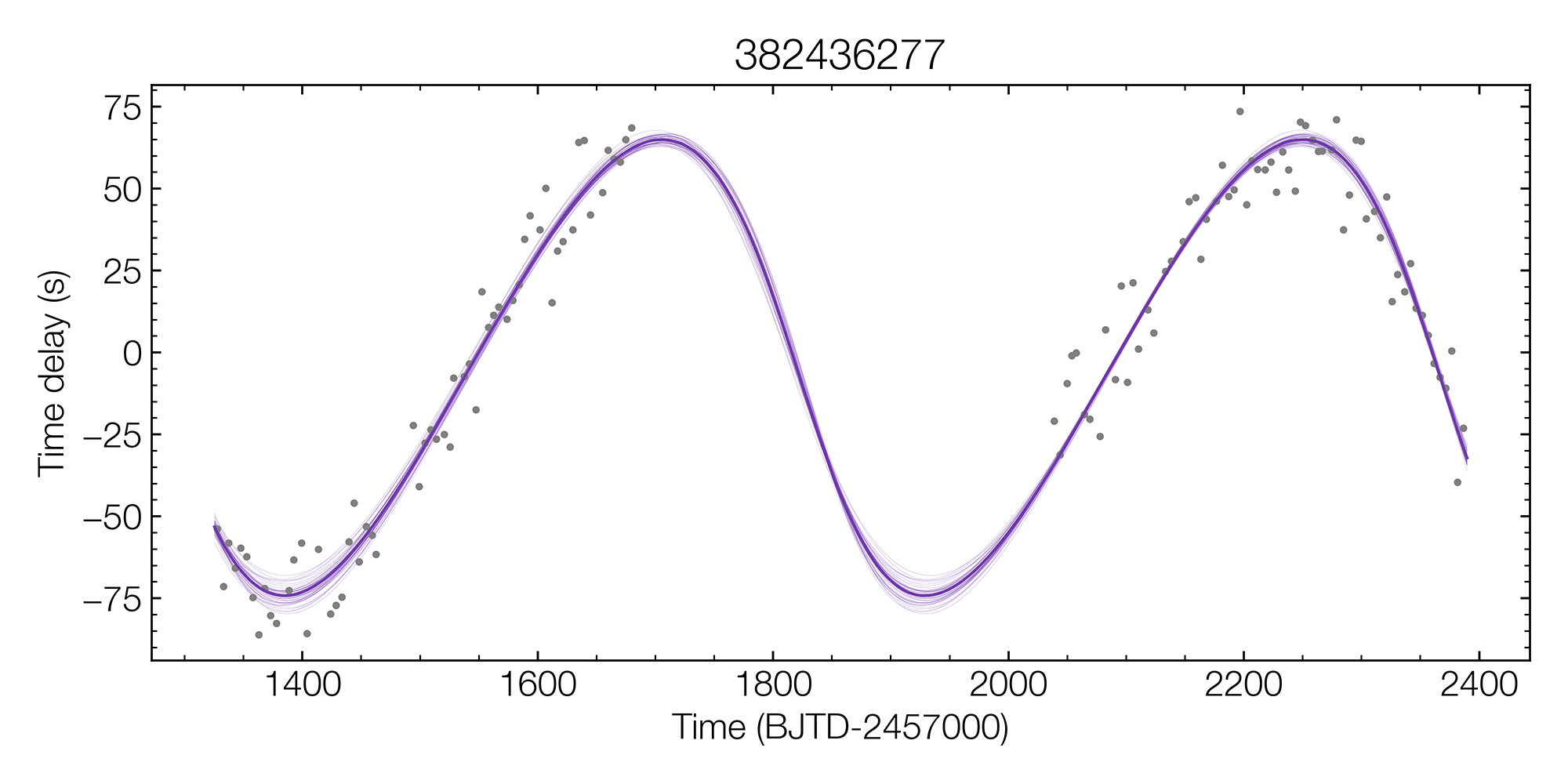}
\includegraphics[width=0.3\textwidth]{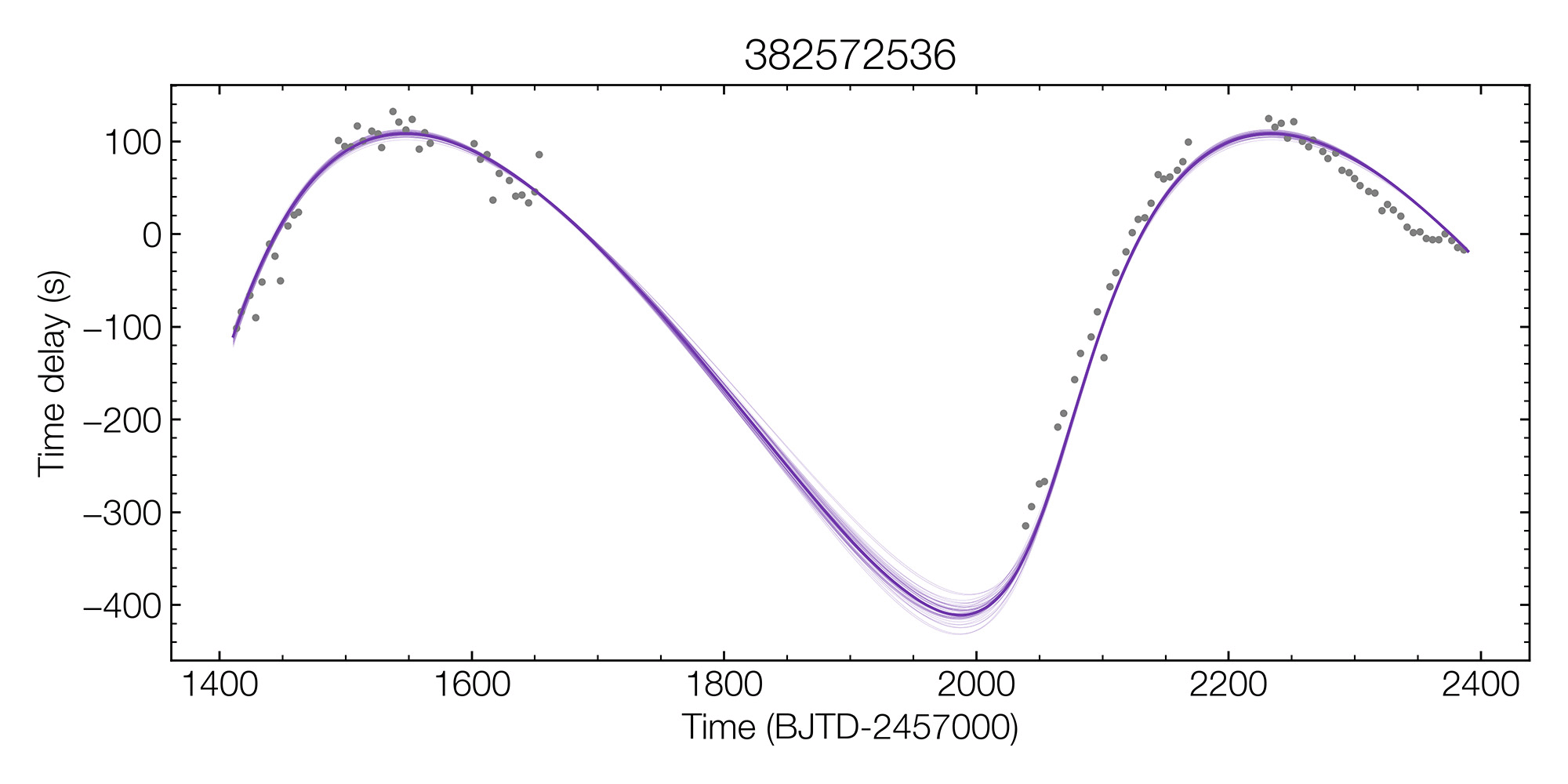}
\includegraphics[width=0.3\textwidth]{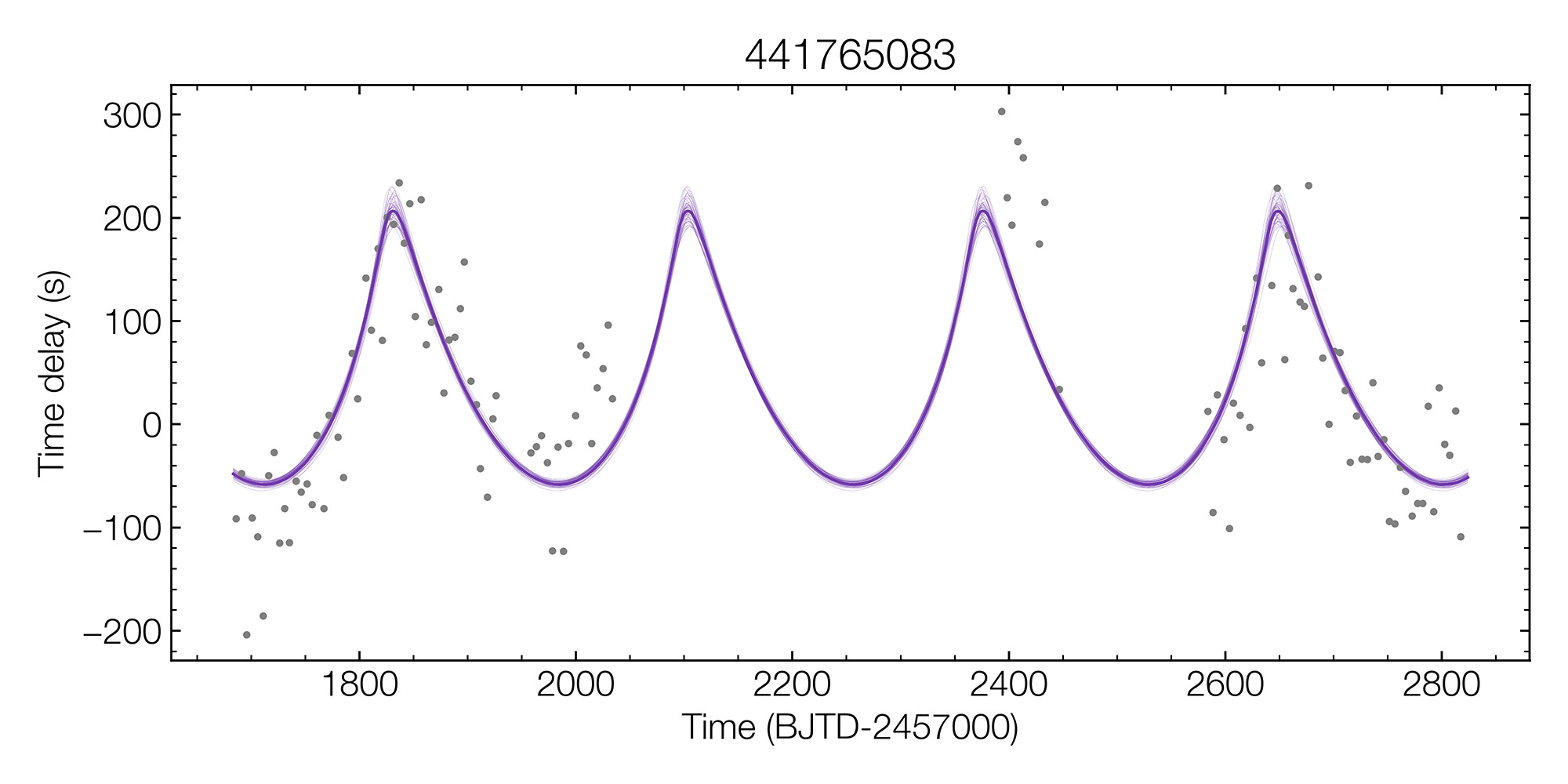}
\includegraphics[width=0.3\textwidth]{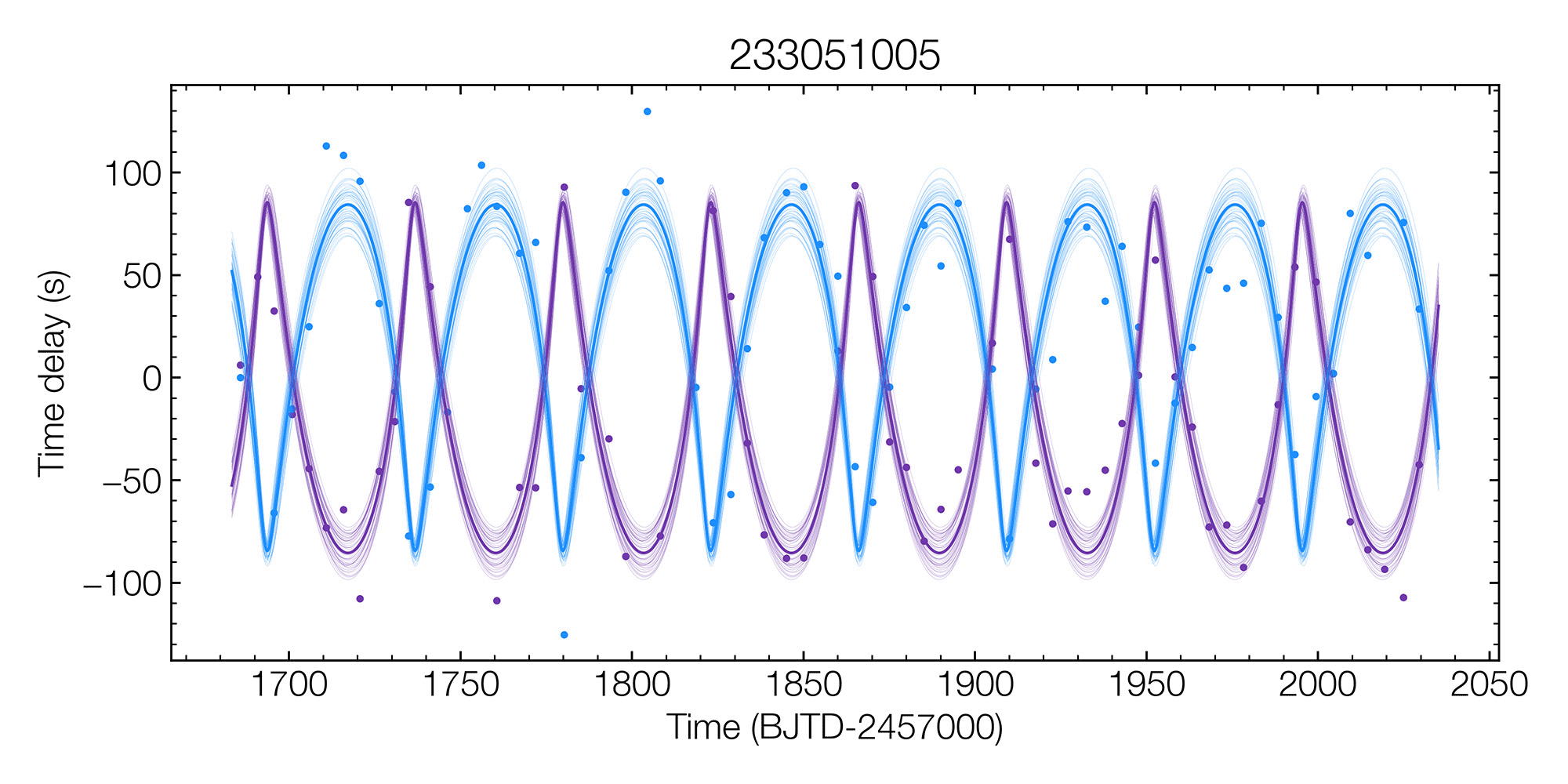}
\caption{Plots of phase modulations of binaries with solvable orbits in \textit{TESS} photometry. Each plot is labeled by TIC ID. The grey points represent the pulsation phases of 5 day windows of the light curves, and the purple lines are the MCMC fit of the phase modulations on the full set of photometry. Random draws from the posterior samples are overplotted in lighter purple to illustrate the uncertainties of the fit. The last plot is the PB2 system TIC 233051005, in which both stars pulsate. The fit to modes from each star are represented in blue and purple respectively.}
\label{fig:mcmcplots}
\end{figure*}

We identify 53 new binaries from our method. For 24 of the binaries we have enough \textit{TESS} data to uniquely determine orbital parameters. These binaries are presented in Table \ref{tab:Binary_summary_table} along with posterior medians and uncertainties for the orbital parameters from our MCMC fit. The fitted phase modulations are presented in Fig.\,\ref{fig:mcmcplots}. The 29 systems for which  the orbital period is longer than the baseline of TESS photometry in our sample are summarized in Table \ref{tab:fullbinaries}. Their long orbital period may result in potentially degenerate fits. Nine of the 53 binaries have astrometric orbital solutions or accelerations described in \textit{Gaia} DR3 \citep{gaia2021}. We detect binaries with mass functions as low as $10^{-3}$ M$_\odot$ and detect periods between 30 and 800\,d. We also present histograms of periods, mass functions, and eccentricities for our binary sample in Fig.\,\ref{fig:hists} with solvable binaries in purple and long-period binaries, which may have degenerate parameters, in grey. 

In our binary sample, a few systems are worthy of note. We identify one target, TIC~233051005, as a PB2 system where both stars pulsate, with the modes of each star exhibiting phase modulation signals of opposite phase. We identify TIC~308396022 as a high-amplitude $\delta$\,Sct variable with a companion in a highly circular orbit, previously published in \cite{yang2021}. We also identify the companion to $\alpha$ Pictoris, which we treat in further detail in Section \ref{sec:alphapic}.

We recover binaries at a range of eccentricities, and recover the occurrence rate trends towards longer periods by \textit{Kepler} in \citet{murphy2018}. This \textit{Kepler} survey had a sample of 2224 $\delta$\,Sct stars and recovered 341 total binaries. Our sample recovers 53 binaries out of 1166 ($\approx 5\%$) stars searched. By comparison, the \textit{Kepler} sample detected 256 binaries with an orbital period shorter than 4 years, the maximum possible baseline of our observations, in a stellar sample of 2224 stars, resulting in a yield of $\approx 11\%$. The lower yield of our sample compared to the \textit{Kepler} can be explained by the presence of significant gaps in the \textit{TESS} timeseries. 

\subsection{$\alpha$ Pictoris}
\label{sec:alphapic}
The brightest star in our sample of phase modulation binaries from \textit{TESS} is the well-known star $\alpha$ Pictoris (V mag = 3.27). The \textit{Hipparcos} satellite strongly suggested a binary companion to $\alpha$ Pictoris due to an astrometric acceleration, but presented a poorly constrained orbit (\citealt{goldin2006},\citealt{vanLeeuwen2007}). \citet{Gullikson2016} identified the likely companion to $\alpha$ Pictoris from spectroscopic data from the CHIRON spectrograph, which they cross correlated to identify the lines of the companion. From this, they suggest a companion with an effective temperature of  $5068 \pm 91$ K and a spectroscopic mass of $0.9 \pm 0.01$~M$_\odot$.  Due to the rapid rotation of $\alpha$ Pictoris (v$\sin$i $\approx$ 200 \,km\,s$^{-1}$), the orbit of the companion remains largely inaccessible through radial velocity observations \citep{Royer2007}.

\begin{figure}
\includegraphics[width = \hsize]{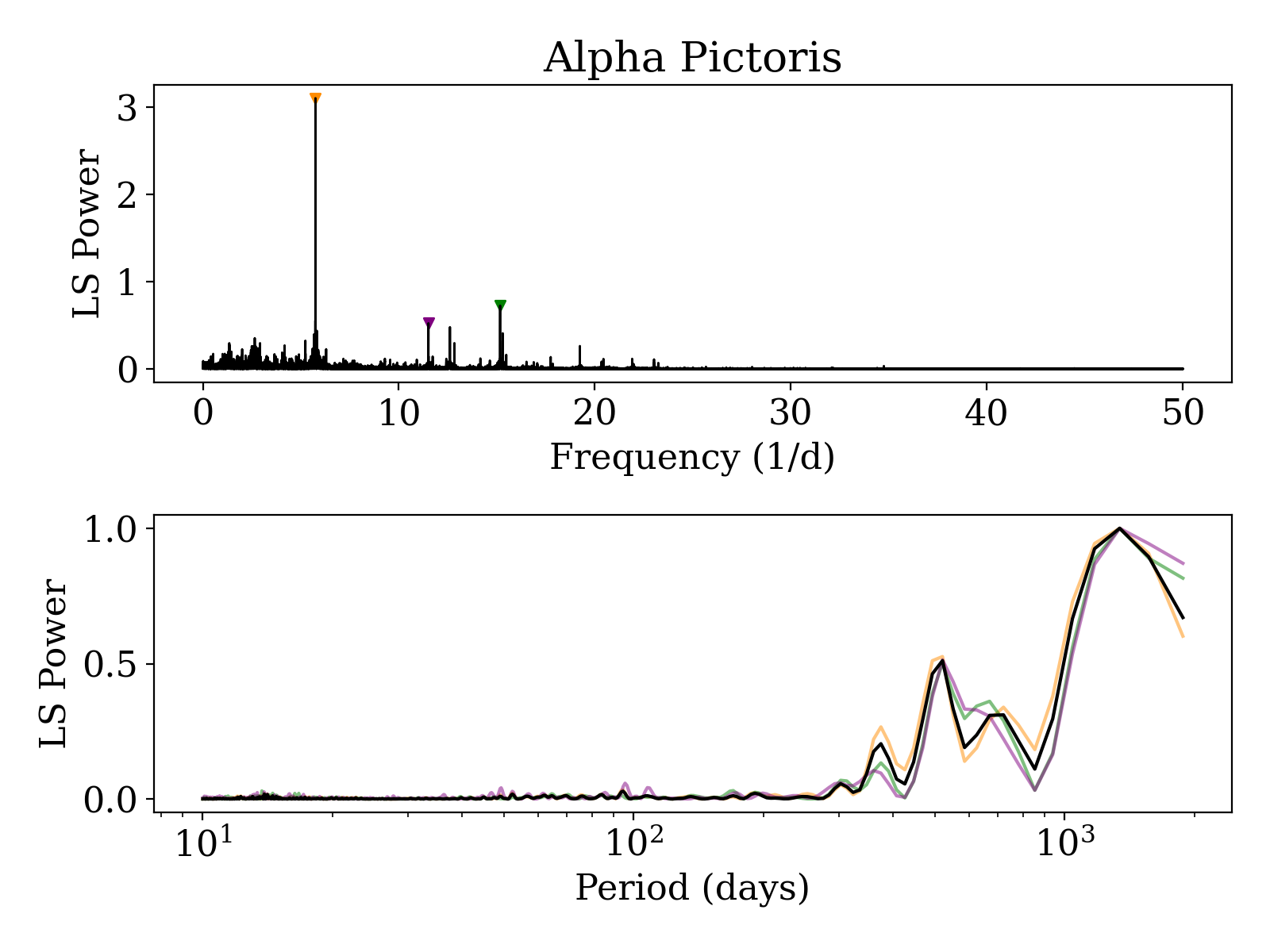}
\caption{Top panel: Lomb Scargle periodogram for $\alpha$ Pictoris, with each mode used in our analysis highlighted by the colored arrow. Bottom panel: Lomb Scargle periodogram of the time delays, with each mode highlighted according to the coloured arrow in the top panel. All modes agree on a periodic phase modulation signal beyond 1000 days.}
\label{fig:alpha_pic_ls}
\end{figure}

We find three $\delta$\,Sct pulsation modes in the \textit{TESS} light curve suitable for analysis. All three modes present a clear, long-period modulation in the pulsation phase. We therefore use the full set of available \textit{TESS} photometry (as of June 2024) until Sector 69 along with the \textit{Hipparcos} astrometry to jointly fit the orbit for this target. 

Our methodology for the model is the same as in Sec. \ref{sec:phasemod} for the fitting of the phase modulations, except that we fit the \textit{Hipparcos} astrometry jointly. We implement the method of \citet{sahlmann2011} in our \texttt{numpyro} model, and make the following modifications. For the TESS phase modulations, we set the deterministic parameter
\begin{eqnarray}
    a_1 \sin(i)/c = \left(\frac{P_{\rm orb}^2 G f_{\rm mass}}{4 \pi ^2}\right)^{(1/3)} / c.
\end{eqnarray} 
where $a_1 \sin(i)/c$ is the light arrival time (with $a_1 \sin(i)$ being the projected semimajor axis of the primary), P$_{\rm orb}$ is the orbital period, and $f_{\rm mass}$ is the mass function. 
As with \citet{sahlmann2011}, we fit a systemic astrometric solution to the \textit{Hipparcos} data with an offset and proper motion in RA and DEC and with a system parallax. We then add a binary star model with the orbital period, primary mass, secondary mass, periastron phase, eccentricity, inclination, and the argument of periapsis, and longitude of node as parameters to the model. We use uniform priors over the domain for all angular parameters and wide lognormal priors on all other parameters centered on the optimization solutions, except for the primary mass, for which we use a normal prior of $1.6 \pm 0.1$~M$_\odot$ from \citet{Gullikson2016}. We additionally multiply a scaling parameter, the intensity ratio $\alpha_{\rm corr}$, to the computed angular semimajor axis from the astrometry. This parameter accounts for the contribution from the flux of both stars in the measured position of the photocenter, which reduces the measured photocenter motion roughly proportional to the flux ratio between the stars in the \textit{Hipparcos} bandpass. We select this value deterministically from the primary and secondary mass based on Eq. 6 of \citet{shahaf2019} and a V magnitude mass luminosity relation, which approximates the \textit{Hipparcos} bandpass well \citep{malkov2007}.

\begin{figure}
\includegraphics[width = \hsize]{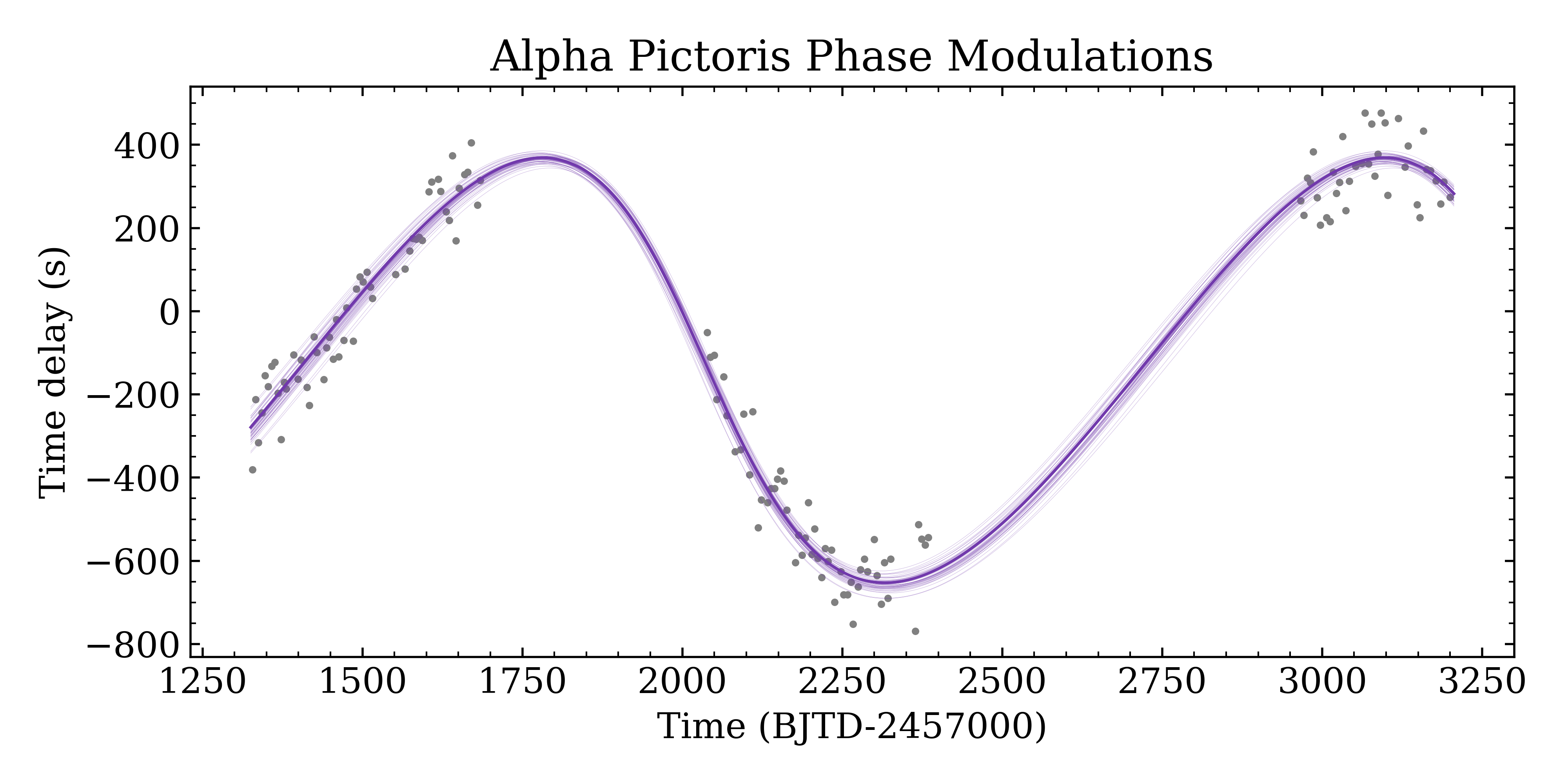}
\caption{Phase modulations of $\alpha$ Pictoris due to the influence of the unresolved binary companion. Grey points are the phase of the pulsation modes in 5-d windows of the light curve. The dark purple line is the median of the samples for the MCMC forward model to the photometry. Fifty random draws from the posterior samples are plotted in light purple to illustrate the uncertainties. We obtain an orbital period for the $\alpha$ Pictoris system of $1316\pm2$\,d.}
\label{fig:alpha_pic}
\end{figure}

We recover the first clear orbit for $\alpha$ Pictoris B (Fig.\,\ref{fig:alpha_pic}), with an orbital period of $1316 \pm 2$\,d and an eccentricity of $0.29 \pm 0.02$. We obtain a companion mass of 1.05~$\pm$~0.05~M$_\odot$, higher than the spectroscopic mass of 0.9$\pm$0.01~M$_\odot$  presented in \citet{Gullikson2016}. This can be explained by an incorrectly estimated spectroscopic primary mass, if the spectroscopic secondary mass uncertainties were underestimated, or if $\alpha$ Pic B is itself a binary. We also obtain an inclination for the system of 121 $\pm$ 2 degrees.

Our fit for $\alpha$ Pictoris also represents the first joint fit of astrometry and pulsation timing, demonstrating the synergy between the two techniques. With the inclusion of astrometric data, we are able to estimate an inclination, allowing for a mass ratio to be estimated as opposed to the mass function with $\sin(i)$ dependence usually obtained from pulsation timing alone. Our procedure can be extended to many of our phase modulation binaries, as well as much of the \textit{Kepler} sample in \citet{murphy2018} with upcoming \textit{Gaia} DR4 astrometry, providing mass ratios even when the precision of the astrometric data is poor.

\begin{figure}
\centering
\includegraphics[width = \hsize]{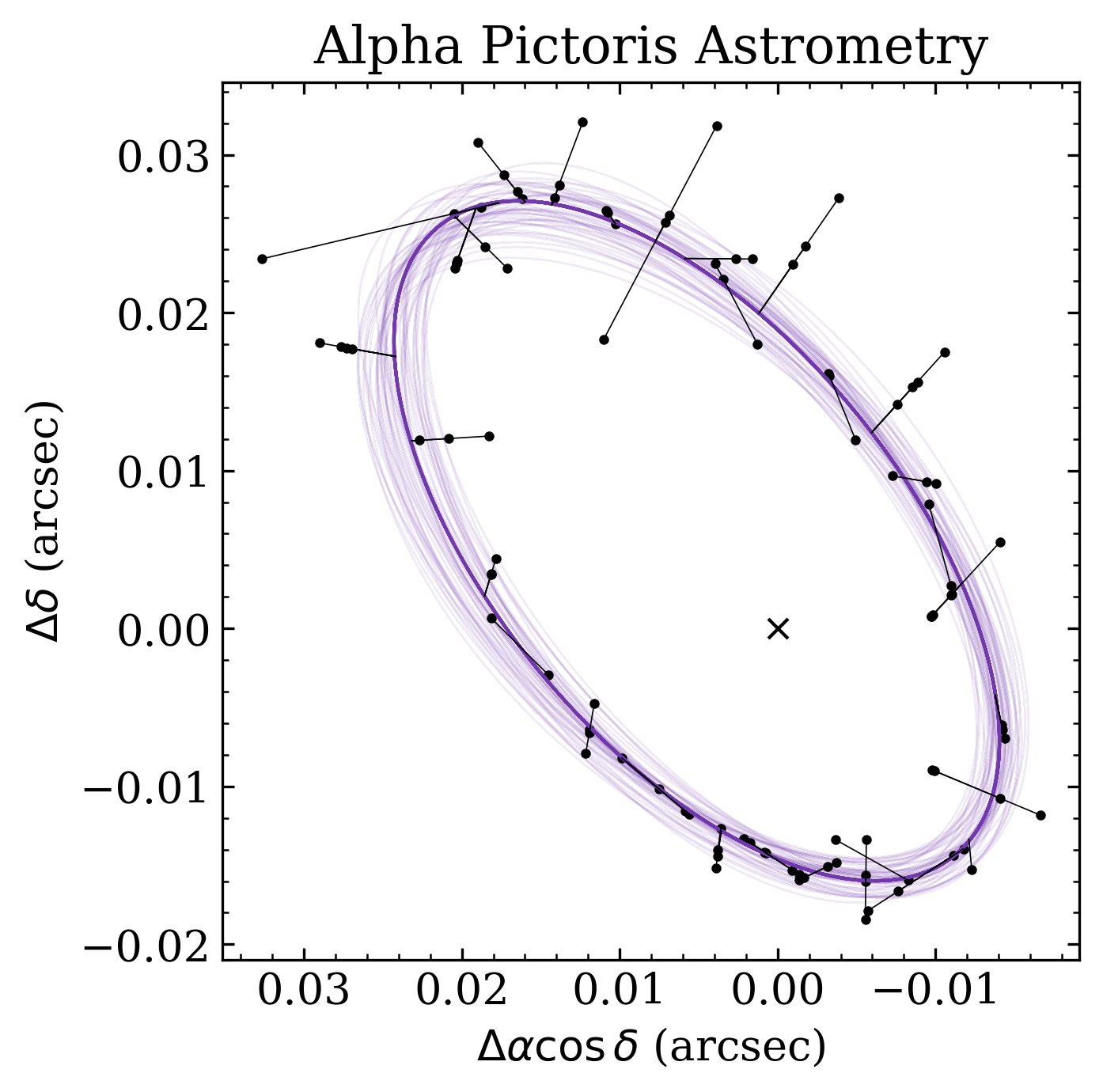}
\caption{\textit{Hipparcos} astrometric photocenter orbit of $\alpha$ Pictoris. The epoch astrometry with the best-fitting single star solution removed is plotted in black points, with black lines drawn towards the best fit model in the direction of the scan angle. The dark purple line shows the best fit model and the light purple lines are draws from the orbital fit to illustrate the uncertainties of the model.}
\label{fig:alpha_pic_astrometry}
\end{figure}

\begin{table}
    \caption{System parameters of $\alpha$ Pictoris.}
    \centering
    \label{tab:star_params}
    \begin{tabular}{lrl}
        \hline\hline
        Parameter & \multicolumn{1}{l}{Value} & Reference \\
        \hline
        RA (J2000) & 06 48 11.4551 & \citet{vanLeeuwen2007} \\
        DEC (J2000) & -61 56 29.0008 & \citet{vanLeeuwen2007} \\
        $\mu_\text{RA}$ (mas) & 70 $\pm$ 1 & This work \\
        $\mu_\text{DEC}$ (mas) & 26 $\pm$ 0.6 & This work \\
        $\Delta$RA & 0.01 $\pm$ 0.003 & This work \\
        $\Delta$DEC & 0.01 $\pm$ 0.002 & This work \\
        V (mag) & 3.27 $\pm$ 0.01 & \citet{hog2000} \\
        \hline
        $T_{\rm eff}$ ($\alpha$ Pic A) & 7770 $\pm$ 264 & \citet{Gullikson2016} \\
        $T_{\rm eff}$ ($\alpha$ Pic B) & 5068 $\pm$ 91 & \citet{Gullikson2016} \\
        Spec. type ($\alpha$ Pic A) & A8V & \citet{Gullikson2016} \\
        Mass ($\alpha$ Pic A) & $1.6\pm{0.1}$ M$_\odot$ & \citet{Gullikson2016} \\
        \hline
        Orbital Period (days) & 1316 $\pm$ 2 & This work \\
        Eccentricity & $0.29 \pm 0.02$ & This work \\
        Inclination (deg) & 121 $\pm$ 2 & This work \\
        Mass ($\alpha$ Pic B; M$_\odot$) & 1.05 $\pm$ 0.05  & This work\\
        Mass Ratio ($\frac{\rm m_2}{\rm m_1}$) & 0.61 $\pm$ 0.04  & This work\\
        $\omega$ (deg) &  353 $\pm$ 4  & This work\\
        $\Omega$ (deg) &  219 $\pm$ 4  & This work\\
        $\phi$ (deg) &  $194 \pm 3$  & This work\\
        t$_{\rm peri}$ (BJD) &  $2456390 \pm 13$  & This work\\
        $\alpha_{\rm corr}$ &  $0.58 \pm 0.02$  & This work\\

        \hline
    \end{tabular}
\end{table}

\subsection{Binary Kinematics}
\label{sec:kinematics}
\citet{murphy2018} suggested that a significant fraction of intermediate-period companions to hot stars are evolved compact objects. In these systems, stable mass transfer through Roche Lobe overflow or efficient wind accretion would be expected in the evolutionary history of the system \citep{Lauterborn1970, McClure1980, Toonen2014}. These binaries are the progenitors of a long list of astrophysically valuable products, including Barium stars, blue stragglers, AM CVn stars, Helium stars, subdwarf stars, R CRb stars, 1991bg-like Type Ia supernovae, and Type 1a supernovae \citep{McClure1990, Preston2000, Solheim2010, Yoon2004, Geier2009, Clayton2012, Pakmor2010, maoz2014}. A secondary consequence of this would be tidal circularization for these systems, a hypothesis that has been supported both by \citep{murphy2018} and by previous studies of binary companions to barium stars (\citealt{jorissen1998}; \citealt{vanWinckel2003}; \citealt{geller2011}).

Kinematics now available from \textit{Gaia} can independently identify candidate field blue stragglers by identifying hot stars with anomalously old kinematic ages. \citet{murphy2018} hypothesized that their observed overabundance of low-eccentricity systems at long periods is consistent with those systems having experienced prior mass transfer. In such cases, the original secondary was of later spectral type and would have been rejuvenated well into its main-sequence life by accreting a significant fraction of its mass from an evolving companion. The original secondary is now the primary and pulsates as a $\delta$\,Sct star, with a rejuvenated lifetime of $\lesssim1$\,Gyr. Depending on the initial mass and configuration of the two stars, such a system's kinematics may be more typical of the thick disc or halo than of the Galactic thin disc.

\begin{figure}
\includegraphics[width = \hsize]{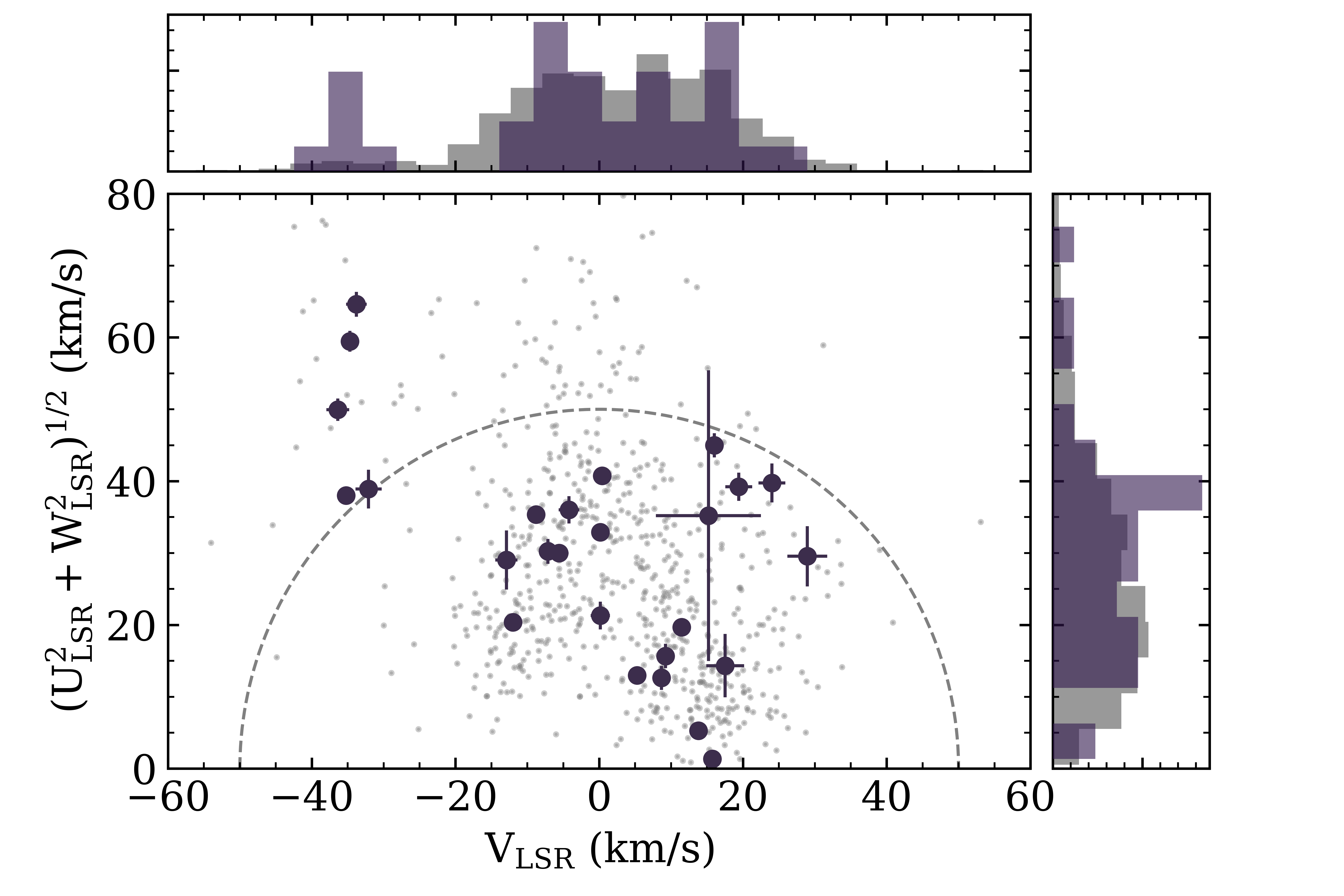}
\caption{Toomre diagram of pulsation modulation binaries from our \textit{TESS} sample. The y-axis is the tangential galactic velocity and the y-axis is the transverse galactic velocity. The dashed semi-circle line marks $50$\,km\,s$^{-1}$. Grey points show pulsators in our sample with no detected phase modulations and purple points represent those in our sample of phase modulation binaries. The histograms show the densities of binaries and stars with no detected phase modulations in purple and grey respectively on each axis. }
\label{fig:toomre}
\end{figure}

\begin{figure}
\includegraphics[width = \hsize]{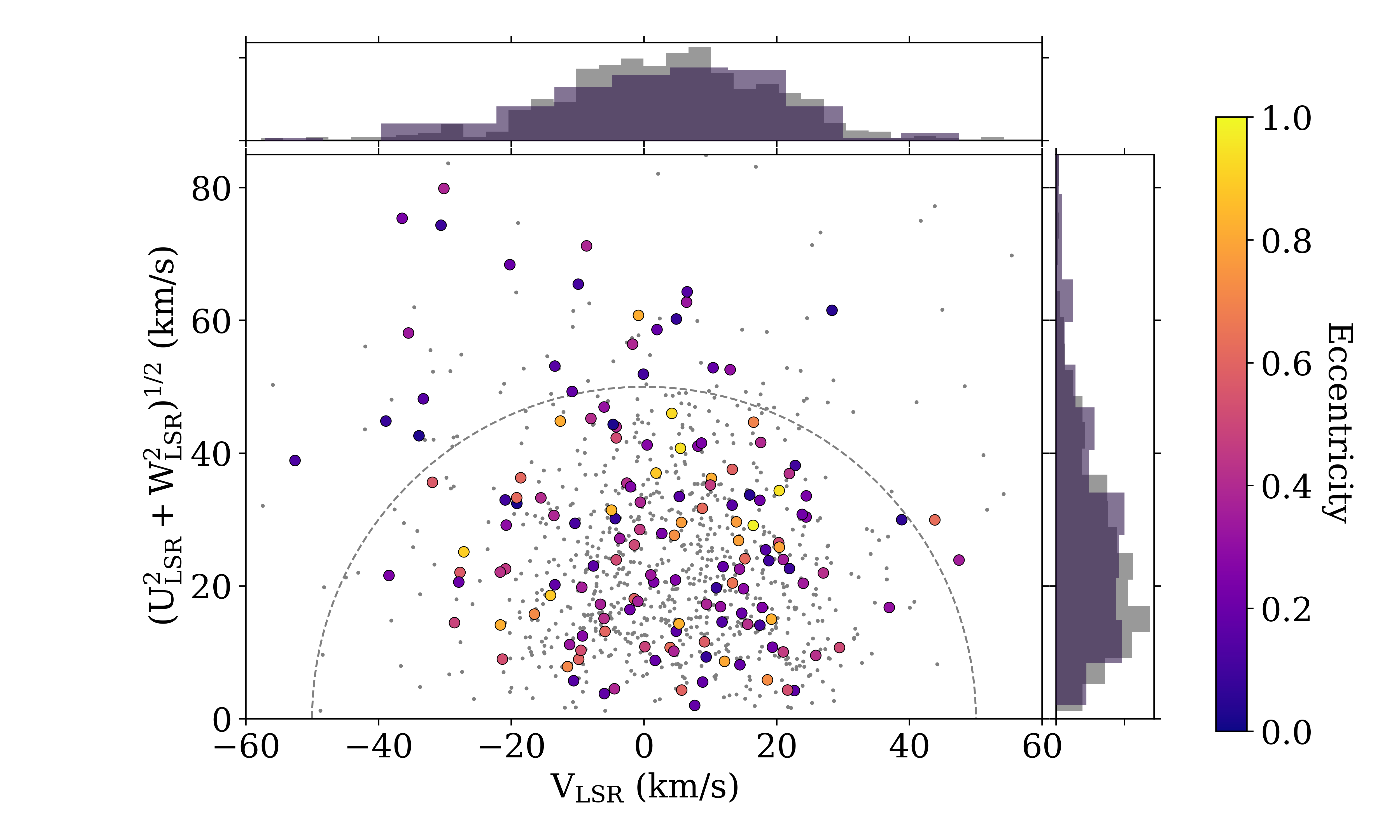}
\caption{Toomre diagram of pulsation modulation binaries from the Kepler sample in \citet{murphy2018}. The coloured points are the phase modulation binaries, colored according to their eccentricity. The y-axis is the transverse component of the galactic velocity relative to the LSR, and the x-axis is the radial component. The dashed semi-circle line marks $50$\,km\,s$^{-1}$. The grey points in the background are stars found in that analysis to be single. The histograms on each axis show the normalized distributions of binaries and single stars in purple and grey respectively.}
\label{fig:kepler_toomre}
\end{figure}

To identify any such systems, we first obtain Galactic kinematics for our sample of binaries. We also revisit the sample of phase modulation binaries from \textit{Kepler} presented in \citet{murphy2018} with the new \textit{Gaia} kinematics. In order to obtain these Galactic kinematics, we transform \textit{Gaia} DR3 proper motions and radial velocities (where available) according to the method of \citet{johnson1987}. We do not analyze targets without proper motions or radial velocities listed in \textit{Gaia} DR3. From this, we plot a Toomre diagram of our sample of binaries and a reference sample of $\delta$\,Sct stars for which we found no evidence of binarity (Fig. \ref{fig:toomre}). We also plot this for the \textit{Kepler} sample (Fig.\,\ref{fig:kepler_toomre}), where the sample size is large enough for us to color-code the binaries by their eccentricity.

In the TESS sample, we see five binaries that appear to lie outside the $50$\,km\,s$^{-1}$ circle, suggestive of stars kinematically hotter than expected for $\delta$\,Sct stars with $\approx$ 2 Gyr main sequence lifetimes. The fraction of the TESS $\delta$\,Sct stars with binary companions outside the circle over the total number of binaries (16 \%) is slightly higher than a similar population fraction for those around which we don't detect binaries (12 \%). Similarly, the fraction of \textit{Kepler} binaries beyond the circle (33.0 \%) is higher than single stars (25.6 \%). Taken together, this tentatively suggests that the population of phase modulation binaries is kinematically hotter than a similar sample of $\delta$\,Sct stars without detectable intermediate-period binary companions.

To calculate more quantitative estimates of the probability of membership in the thick disk, we employ the method of \citet{bensby2003}. We estimate ${P_{\rm TD}}/{P_{\rm D}}$ and ${P_{\rm H}}/{P_{\rm TD}}$, which are the ratio of probability of Thick-Disk (TD) membership to thin-Disk (D) membership and the ratio of probability of Halo (H) membership to thick disk. We use 
\begin{eqnarray}
    \frac{P_{\rm TD}}{P_{\rm D}} = \frac{X_{\rm TD}}{X_{\rm D}} \frac{f_{\rm TD}}{f_{\rm D}}
\end{eqnarray}
and 
\begin{eqnarray}
    \frac{P_{\rm TD}}{P_{\rm H}} = \frac{X_{\rm TD}}{X_{\rm H}} \frac{f_{\rm TD}}{f_{\rm H}},
\end{eqnarray} 
where $X_{\rm TD}$, $X_{\rm D}$, and $X_{\rm H}$ are the \textit{a priori} probabilities of a star in the Solar neighborhood being in the thick disk, thin disk, and halo respectively (given in \citealt{bensby2003}), and $f_{\rm TD}$, $f_{\rm D}$, and $f_{\rm H}$ are the relative probabilities of a given star being within the respective kinematic distributions of the thick and thin disk respectively, assuming these distributions are Gaussian. These are given by:
\begin{eqnarray}
f = \exp \left( - \frac{U_{\rm LSR}^2}{2 \sigma_{\rm U}^2} - \frac{(V_{\rm LSR} - V_{\rm Asym})^2}{2 \sigma_{\rm V}^2} - \frac{W_{\rm LSR}^2}{2 \sigma_{\rm W}^2}\right),
\end{eqnarray}
where $U_{\rm LSR}^2$,$V_{\rm LSR}^2$,$W_{\rm LSR}^2$ are galactic velocities of a given target. The parameters $\sigma_{\rm U}^2$, $\sigma_{\rm V}^2$, and $\sigma_{\rm W}^2$ are the kinematic dispersions of the stellar population and $V_{\rm Aysm}$ is the asymmetric drift for the population, given for the thick disk, thin disk, and halo in \citet{bensby2003}.

The thick disk is estimated to be 10\,Gyr old \citep{helmi2018} and the halo 10-13\,Gyr, contradicting the approximately 1--2\,Gyr main-sequence lifetime of the $\delta$\,Sct stars. Thus, any systems in these populations are probably field blue stragglers.

We identify no binaries in our \textit{TESS} sample consistent with thick-disk or halo membership. We identify three stars in the \textit{Kepler} sample of binaries which are likely thick-disk members with ${P_{\rm TD}}/{P_{\rm D}}>1$: KIC~4243984, KIC~6861400, and KIC~8500899. In addition, KIC\,11754974, previously identified as a candidate blue straggler in \citet{murphy2013}, exhibits kinematics consistent with the Galactic halo (${P_{\rm TD}}/{P_{\rm H}}<1$), with a tangential velocity relative to the standard of rest of $322 \pm 4$\,km\,s$^{-1}$ and a radial component of $- 388 \pm 7$\,km\,s$^{-1}$. KIC\,11754974 also exhibits a very circular orbit, with an eccentricity of 0.01. Further follow up of these systems, especially with \textit{Gaia} DR4 astrometry, which can identify the inclination of the orbits, may be able to test whether the companions are likely to be white dwarfs as expected. A majority of systems outside the $50$\,km\,s$^{-1}$ circle in Fig.\,\ref{fig:kepler_toomre} are consistent with low-eccentricity orbits, suggesting a history of mass-transfer even for stars that are not confidently members of the thick-disk. The eccentricities of \textit{Kepler} binaries beyond the thin disk are all low (< 0.1). In contrast, 10.4 \% of the \textit{Kepler} binaries in the thin disk have eccentricities lower than 0.1.

\section{Conclusions}
\label{sec:concl}
We searched for pulsation modulation caused by binary companions to 1166 pulsating $\delta$\,Sct stars in TESS photometry from the first four years of its operation. We find 53 binaries in our sample and fit the time delays of 24 binaries, obtaining orbital parameters and mass functions. From their Galactic kinematics, we find that all of these belong to the Galactic thin disc. We also identify four \textit{Kepler} phase modulation binaries from \citet{murphy2018} that belong to the thick disc or halo that are probably field blue stragglers. These systems would have been rejuvenated by mass transfer from a companion which is now a white dwarf. We also measure orbital parameters for the $\alpha$ Pictoris system for the first time, from both pulsational phase modulations in \textit{TESS} data and \textit{Hipparcos} astrometry.

\section{Acknowledgements}

CXH is supported by Australian Research Council (ARC) DECRA project DE200101840. SJM is also supported by the ARC through Future Fellowship FT210100485. We thank Daniel Hey and Jiayin Dong for valuable feedback and advice on the methodology. Funding for the \textit{TESS} mission is provided by NASA’s Science Mission directorate. We acknowledge the use of public \textit{TESS} Alert data from pipelines at the \textit{TESS} Science Office and at the \textit{TESS} Science Processing Operations Center.

\section{Data Availability}

This paper includes data collected by the \textit{TESS} mission, which are publicly available from the Mikulski Archive for Space Telescopes (MAST). Data from the \textit{Hipparcos} spacecraft are also used, which are available in \citet{vanLeeuwen2007} and are available online at https://www.cosmos.esa.int/web/hipparcos/hipparcos-2.

\begin{landscape}
\begin{table}
    \centering
    \caption{Orbital and stellar parameters for the solvable phase modulation binaries in our sample. }
    \begin{tabular}{rrrrrrr@{ $\pm$ }lr@{ $\pm$ }lr@{ $\pm$ }lr@{ $\pm$ }l}
    %\begin{tabular}{rr@{ $\pm$ }lr@{ $\pm$ }llr@{ $\pm$ }lcrrrrrrrll}
        \hline
        \hline
                \multicolumn{1}{c}{ID} &  TESS T & RUWE & $T_{\rm eff}$ & \multicolumn{1}{c}{RA} & \multicolumn{1}{c}{DEC} & \multicolumn{2}{c}{Period} & \multicolumn{2}{c}{Eccentricity} & \multicolumn{2}{c}{Mass func} &  \multicolumn{2}{c}{a $\sin(i)/c$ } \\

      & (mag) &  & (K) & (deg) & (deg) & \multicolumn{2}{c}{(days)} & \multicolumn{2}{c}{} & \multicolumn{2}{c}{(M$_\odot$)} &  \multicolumn{2}{c}{(s)} \\
\hline
219810122&9.08& 4.91 & 7583 &266.9792&64.0984&398&3&0.82&0.02&0.13&0.03&270&20 \\
364323133&7.75& 3.51 & 7657 &87.8793&-63.5312&399.6&0.8&0.42&0.02&0.0064&0.0003&98&1 \\
258351350&7.94& 1.48 & 8015 &296.5106&69.9192&771&4&0.8&0.03&0.062&0.006&320&10 \\
272089183&8.97& ~ & 6698 &115.4068&-75.7673&299&1&0.245&0.009&0.247&0.004&274&1 \\
30316134&11.07& 1.07 & 8550 &74.3834&-68.0191&222&1&0.05&0.02&0.0049&0.0003&61&1 \\
41258067&11.75& 1.91 & 7950 &91.8808&-71.0378&367.8&0.8&0.32&0.01&0.0227&0.0002&141.9&0.3 \\
30537037&10.26& 0.82 & 6340 &75.3296&-71.4204&398&1&0.06&0.02&0.03&0.002&163&3 \\
274674793&7.98& 1.38 & 7768 &256.1068&52.8185&61.68&0.01&0.5&0.01&0.094&0.003&69.3&0.7 \\
349835304&11.12& 1.23 & 7462 &113.8746&-63.8161&105.8&0.4&0.67&0.05&0.008&0.002&44&3 \\
160582982&7.9& 1.18 & 8927 &242.6556&72.3932&89.27&0.05&0.55&0.01&0.417&0.007&145.7&0.9 \\
229945862&9.34& 0.87 & 8644 &288.1967&64.177&114.5&0.3&0.07&0.03&0.4&0.02&170&3 \\
260353074&8.97& 3.71 & 7229 &94.836&-54.8656&200.29&0.05&0.56&0.006&0.512&0.008&268&1 \\
261061652&8.24& 2.29 & 7348 &81.1477&-78.3443&226&2&0.28&0.02&0.053&0.003&136&2 \\
441765083&9.6& 2.75 & 8066 &261.1454&77.156&272.6&0.3&0.67&0.02&0.038&0.002&138&2 \\
279613634&8.78& 5.90 & 7354 &106.4419&-56.2223&503&1&0.446&0.008&0.0245&0.0007&179&2 \\
230067654&11.0& 1.06 & 7543 &242.4243&67.6901&36.36&0.03&0.17&0.03&0.059&0.003&41.8&0.7 \\
382436277&8.16& 2.31 & 7970 &117.1425&-64.8226&544.1&0.8&0.28&0.01&0.00139&4e-05&72.6&0.6 \\
382258769&10.66& 1.01 & 7530 &82.5664&-54.6857&57.92&0.08&0.41&0.04&0.21&0.02&87&2 \\
235709086&6.72& 3.69 & 7599 &289.7333&73.5509&209.0&0.2&0.116&0.005&0.0818&0.0006&149.3&0.4 \\
382572536&9.24& 2.18 & 9047 &118.3123&-64.6984&686.0&0.4&0.494&0.005&0.0567&0.0009&292&2 \\
233051005&7.22& 0.91 & 8092 &268.9292&61.3959&43.1&0.05&0.65&0.04&0.38&0.04&86&3 \\
306773428&7.47& 3.59 & 7975 &121.1308&-67.2082&188.28&0.09&0.897&0.002&0.0124&0.0009&74&2 \\
219776919&9.97& 1.98 & 6643 &263.668&60.3229&185&1&0.75&0.05&0.17&0.05&170&20 \\
308396022&11.07& 2.24 & 7371 &120.2599&-63.6751&804.0&0.6&0.07&0.01&0.024&0.001&243&4 \\

        \end{tabular}
    \label{tab:Binary_summary_table}
    \end{table}

\end{landscape}

\newpage

\newpage
\clearpage
\clearpage
\clearpage
\onecolumn
\begin{table}
\caption{Parameters for stars with observed phase modulations but not labeled as `Solved Binaries'. Orbital parameters and uncertainties from an MCMC fit are listed where possible, but should be used with caution. For orbital parameters of TIC 167602316, see Table \ref{tab:star_params}.}
\label{tab:fullbinaries}
\centering

\begin{tabular}{rrrrrrrr@{ $\pm$ }lr@{ $\pm$ }llr@{ $\pm$ }l}
\hline
\hline
    \multicolumn{1}{c}{ID} & Disposition & TESS T & RUWE & $T_{\rm eff}$ & \multicolumn{1}{c}{RA} & \multicolumn{1}{c}{DEC} & \multicolumn{2}{c}{Period} & \multicolumn{2}{c}{Eccentricity} & Mass func &  \multicolumn{2}{c}{a $\sin(i)/c$ } \\

    &  & (mag) &  & (K) & (deg) & (deg) & \multicolumn{2}{c}{(days)} & \multicolumn{2}{c}{} & (M$_\odot$) &  \multicolumn{2}{c}{(s)} \\
\hline
167602316 & Long-period Binary & 2.89 & ~ & 7510 & 102.05 & -61.94 \\
229954300 & Insufficient Data & 7.48 & 1.24 & 9846 & 288.68 & 65.27 \\
141865875 & Long-period Binary & 7.69 & 1.16 & 7365 & 93.8 & -77.05 & 2000 & 200 & 0.17 & 0.04 & 0.01 & 320 & 30 \\
219861551 & Long-period Binary & 8.03 & 10.52 & 7560 & 258.16 & 67.62 & 1350 & 50 & 0.24 & 0.04 & 0.012 & 270 & 10 \\
233580734 & Long-period Binary & 8.05 & 6.36 & 7905 & 282.65 & 61.98 & 2230 & 10 & 0.021 & 0.001 & 0.015 & 414 & 3 \\
300447314 & Long-period Binary & 8.07 & 3.01 & 7837 & 111.97 & -66.98 \\
308994213 & Long-period Binary & 8.15 & ~ & 8260 & 123.21 & -63.98 \\
256978045 & Long-period Binary & 8.26 & 1.01 & 8641 & 302.29 & 64.05 & 530 & 10 & 0.3 & 0.1 & 0.42 & 480& 40 \\
177082055 & Long-period Binary & 8.28 & 12.0 & 7301 & 103.55 & -68.16 & 1800 & 100 & 0.14 & 0.04 & 0.014 & 350 & 60 \\
377190658 & Long-period Binary & 8.32 & 0.94 & 9614 & 285.78 & 59.44 & 526.1 & 0.7 & 0.32 & 0.01 & 0.12 & 316 & 2 \\
259130275 & Long-period Binary & 8.33 & 6.52 & 7210 & 291.91 & 68.65 & 2450 & 50 & 0.512 & 0.006 & 0.054 & 670 & 10 \\
350622144 & Long-period Binary & 8.83 & 5.17 & 7896 & 87.68 & -56.91 \\
388129392 & Long-period Binary & 8.93 & 1.22 & 7973 & 63.57 & -67.11 \\
353853600 & Long-period Binary & 9.5 & 1.16 & 7758 & 271.17 & 59.66 & 2700 & 30 & 0.580 & 0.008 & 0.088 & 840 & 10 \\
289572269 & Long-period Binary & 9.87 & 2.74 & 7428 & 252.12 & 61.96 \\
229400643 & Long-period Binary & 9.9 & 0.96 & 7526 & 292.49 & 61.75 & 507 & 2 & 0.47 & 0.04 & 0.028 & 189 & 5 \\
229770586 & Long-period Binary & 9.93 & 4.4 & 7866 & 282.37 & 69.56 & 1230 & 40 & 0.09 & 0.03 & 0.048 & 410 & 10 \\
376998123 & Long-period Binary & 9.95 & 2.77 & 7678 & 284.65 & 56.86 & 1200 & 10 & 0.10 & 0.01 & 0.039 & 373 & 5\\
420114772 & Long-period Binary & 10.05 & 7.72 & 7015 & 290.43 & 74.76 & 522 & 3 & 0.45 & 0.06 & 0.00034 & 44 & 1 \\
258827801 & Long-period Binary & 10.31 & 5.05 & 7787 & 287.42 & 67.91 & 532.3 & 0.4 & 0.898 & 0.001 & 0.52 & 514 & 2 \\
288241831 & Long-period Binary & 10.63 & 5.27 & 7690 & 258.79 & 78.21 \\
177240597 & Long-period Binary & 11.01 & 1.7 & 7101 & 101.3 & -77.03 & 283 & 3 & 0.62 & 0.06 & 0.13 & 215 & 9 \\
179638690 & Long-period Binary & 11.06 & 1.01 & 9326 & 80.6 & -67.37 \\
141865028 & Long-period Binary & 11.39 & 2.77 & 7257 & 93.52 & -76.73 & 356 & 5 & 0.21 & 0.03 & 0.044 & 173 & 4 \\
141684647 & Long-period Binary & 11.49 & 3.97 & 7448 & 90.46 & -72.74 & 413 & 5 & 0.15 & 0.01 & 0.11 & 256 & 5 \\
142050276 & Long-period Binary & 11.6 & 1.41 & 7991 & 97.57 & -73.96 & 299 & 2 & 0.56 & 0.02 & 0.044 & 154 & 1 \\

\end{tabular}
\end{table}
\clearpage
\newpage
\twocolumn

\bibliographystyle{mnras}
\bibliography{references} % if your bibtex file is called example.bib
\end{document}